%% file: MAIN.tex
\documentclass[11pt]{article}
\usepackage[a4paper,hmargin=1.0in,vmargin=1.0in]{geometry}
\usepackage{amsmath,amsthm,amssymb,setspace,sectsty,bbm,graphicx,mathtools,thmtools,subcaption}
\usepackage[round]{natbib}
\usepackage[linktocpage=true,pagebackref=true,colorlinks,allcolors=blue,bookmarks=true]{hyperref}
\usepackage[dvipsnames]{xcolor}

\newtheoremstyle{DStheorem}% name of the style to be used
{\topsep}% measure of space to leave above the theorem. E.g.: 3pt
{\topsep}% measure of space to leave below the theorem. E.g.: 3pt
{\itshape}% name of font to use in the body of the theorem
{0pt}% measure of space to indent
{\scshape}% name of head font
{.}% punctuation between head and body
{ }% space after theorem head; " " = normal interword space
{\thmname{#1}\thmnumber{ #2}\thmnote{ (#3)}}
\theoremstyle{DStheorem}
\newtheorem{theorem}{Theorem}[section]
\newtheorem{lemma}[theorem]{Lemma}
\newtheorem{claim}[theorem]{Claim}

\let\oldproofname=\proofname
\renewcommand{\proofname}{\rm\sc{\oldproofname}}

\newcommand{\bs}[1]{\boldsymbol{#1}}
\newcommand{\bstitle}[1]{\texorpdfstring{$\boldsymbol{#1}$}{}}
\newcommand{\eps}{\epsilon}

\newcommand{\expar}[1]{{\mathbb E} [ #1 ]}

\newcommand{\opt}{\mathrm{OPT}}

\newcommand{\Tearly}{{\cal T}_{\mathrm{early}}}
\newcommand{\Tmid}{{\cal T}_{\mathrm{mid}}}
\newcommand{\Tlate}{{\cal T}_{\mathrm{late}}}
\newcommand{\myheavy}{\mathrm{heavy}}
\newcommand{\mylight}{\mathrm{light}}

\sectionfont{\large} \subsectionfont{\normalsize}
\allowdisplaybreaks
\makeindex
\newcommand{\changelocaltocdepth}[1]{%
  \addtocontents{toc}{\protect\setcounter{tocdepth}{#1}}%
  \setcounter{tocdepth}{#1}%
}

\newcounter{gidicounter}
\newcounter{dannycounter}

\begin{document}

\begin{titlepage}

\title{Improved Approximation Guarantees and Hardness Results for \\
MNL-Driven Product Ranking}
\author{%
Danny Segev\thanks{School of Mathematical Sciences and Coller School of Management, Tel Aviv University, Tel Aviv 69978, Israel. Email: {\tt segevdanny@tauex.tau.ac.il}. Supported by Israel Science Foundation grant 1407/20.}%
\and%
Gidi Steinberg\thanks{School of Mathematical Sciences, Tel Aviv University, Tel Aviv 69978, Israel. Email: {\tt gidisteinberg@gmail.com}.}}
\date{}
\maketitle

\pagenumbering{Roman}

\begin{abstract}
In this paper, we address open computational questions regarding the market share ranking problem, recently introduced by \cite{DerakhshanGMM22}. Their modelling framework incorporates the extremely popular Multinomial Logit (MNL) choice model, along with a novel search-based consider-then-choose paradigm. In a nutshell, the authors devised a Pandora's-Box-type search model, where different customer segments sequentially screen through a ranked list of products, one position after the other, forming their consideration set by including all products viewed up until terminating their inspection procedure. Subsequently, a purchasing decision out of this set is made based on a joint MNL choice model.

Our main contribution consists in devising a polynomial-time approximation scheme for the market share ranking problem, utilizing fresh technical developments and analytical ideas, in conjunction with revising the original insights of  \cite{DerakhshanGMM22}. Along the way, we introduce a black-box reduction, mapping general instances of the market share ranking problem into ``bounded ratio'' instances, showing that this result directly leads to an elegant and easily-implementable quasi-PTAS. Finally, to provide a complete computational characterization, we prove that the market share ranking problem is strongly $\mathrm{NP}$-hard. 
\end{abstract}


\end{titlepage}

% CONTENTS %%%%%%%%%%%%%%%%%%%%%%%%%%%%%%%%%%%%%%%
\newpage
\setcounter{page}{2}
\tableofcontents

% SECTIONS %%%%%%%%%%%%%%%%%%%%%%%%%%%%%%%%%%%%%%%%%%%%%%%%%%%%%%%%%%%
\newpage
\pagestyle{plain}
\pagenumbering{arabic}
\setcounter{page}{1}

\input{TEX-Intro}
\input{TEX-Hardness}
\input{TEX-QPTAS.tex}
\input{TEX-PTAS-Overview}
\input{TEX-Partition}
\input{TEX-Conclusions}

% BIB %%%%%%%%%%%%%%%%%%%%%%%%%%%%%%%%%%%%%%%%%%%
\addcontentsline{toc}{section}{Bibliography}
\bibliographystyle{plainnat}
\bibliography{BIB-Ranking}

% APPENDIX %%%%%%%%%%%%%%%%%%%%%%%%%%%%%%%%%%%%%%
\changelocaltocdepth{1}
\appendix
\input{TEX-Appendix}

\end{document}

%% file: TEX-Intro.tex
\section{Introduction}

In the last few decades, assortment optimization has been extensively studied in a broad range of domains spanning among others online retail, e-commerce, brick and mortar, marketing, and advertising. In such settings, retailers typically seek to enhance various performance metrics by offering consumers the ``optimal'' collection of products. Technically speaking, given an underlying collection of products, we wish to determine its best-possible subset, in the sense of maximizing revenue, appealing to the widest customer segment, or meeting additional structural criteria. Such problems often revolve around ``choice models'', i.e., mathematical depictions of presumed customer behavior, aiming to rigorously capture the decision outcomes of consumers contemplating between numerous purchasing alternatives. In these scenarios, we are facing an inherent trade-off between the prediction power of a given choice model in capturing real-life behavioral phenomena and its computational tractability in various operational settings, both from a parameter estimation standpoint as well as from an optimization perspective. In light of these complexities, problems of this nature have been receiving a great deal of attention from academics, consulting firms, and software companies alike, opting to advance the theoretical foundations of this field while concurrently focusing on concrete market solutions. For a detailed literature review on this topic, readers are referred to the excellent books of \cite{BookGallegoTopaloglu2019} and \cite{Phillips2021}, as well as to the references therein.

\paragraph{Product ranking problems.} Building on the classical setting described above, which can be viewed as a ``subset selection'' problem, we have recently witnessed novel modeling frameworks, crafted to tackle the real-life intricacies of modern assortment optimization. Such approaches include, for example, dynamic assortment planning, network revenue management, multi-purchase models, and display optimization. Specifically related to the last direction, in order to better suit modern-day applications, product ranking problems have recently been a very active domain. Here, we take into account the relative positions in which products are located along with the effect of these locational decisions on their exposure to prospective buyers, leading to a more fine-grained modeling approach. This research direction has rapidly evolved in the last few years, as one can discover by consulting selected papers along these lines \citep{FerreiraParthasarathy21, asadpour2022sequential, CompianiLewisPengWang21, golrezaei2020learning, agarwal2024misalignment}. 

\paragraph{Search-based formation of consideration sets.}\label{search_then_cons} As previously mentioned, the vast majority of product ranking problems are centered around choice models, which serve as a functional representation of the way customers make decisions when facing multiple purchasing alternatives. This setting typically consists of two stages, where each customer initially browses through a ranked list of alternatives, forming her so-called consideration set, and then makes a purchasing decision out of this set via a given choice model. For an additional background on such approaches, we point readers to the work of \cite{FeldmanTopaloglu}, \cite{AouadFariasLeviSegev18},  \cite{FloresBerbegliaVanHentenryck2019}, \cite{AouadFariasLevi2020}, \cite{AouadS21}, and \cite{SegevFeldman22}.

In this paper, we address open computational questions regarding the market share ranking problem, recently introduced by \cite{DerakhshanGMM22}. Their modelling framework incorporates the extremely popular Multinomial Logit (MNL) choice model, along with a novel search-based consider-then-choose paradigm. In a nutshell, \cite{DerakhshanGMM22} devised a Pandora's-Box-type search model \citep{Weitzman1979}, where different customer segments sequentially screen through a ranked list of products, one position after the other, forming their consideration set by including all products viewed up until terminating their inspection procedure. Subsequently, a purchasing decision out of this set is made based on a joint MNL choice model.

In what follows, we succinctly describe the above-mentioned search process, putting a special emphasis on its random utility structure and customers' welfare function. The basic building blocks of this process are $n$ substitutable products, where the random utility of each product $i \in [n]$ takes the form $U_i + Z_i$. Here, $U_i$ is a random variable representing the intrinsic utility of a yet-unobserved product, and $Z_i \sim \mathrm{Gumbel}(0,1)$ is an idiosyncratic shock. For convenience, we make use of $W_i = e^{U_i}$ to denote the so-called preference weight of product $i$, with the assumption that $U_1, \ldots, U_n$ are independent and identically distributed; the same goes for the random shocks $Z_1, \ldots, Z_n$. Now, the combinatorial structure through which our search evolves is a linear order, determined by a position-to-product assignment $\mathcal{A} : [n] \to [n]$, where $\mathcal{A}(p)$ designates the product placed in position $p$ of this ordering. Given this assignment, a representative customer sequentially inspects products, one position after the other, deciding at each position $p$ whether to terminate her search procedure, yielding the consideration set $\mathcal{A}(1), \dots, \mathcal{A}(p)$, or to inspect the next product $\mathcal{A}(p+1)$ at a cost of $c_{p+1}$, which depends only on its position. This procedure continues up until actively deciding to stop at some position or exhausting the entire collection of products, in which case the resulting consideration set naturally consists of all products. Along these dynamics, having inspected the prefix of positions $1, \ldots, p$, the customer's welfare function is defined as  
\[
\mathrm{Welfare}(p) ~~=~~ \mathbb{E}\!\left[\max_{q \in [p]}\left\{\ln (W_{\mathcal{A}(q)}) + Z_{\mathcal{A}(q)}\right\}\right] - \sum_{q \in [p]} c_q \ , 
\] 
with the first term being the expected maximum utility across all inspected products, and the second term standing for the total search cost associated with positions $1, \ldots, p$. In this setting, the customer's objective is to devise a stopping time that maximizes her expected welfare.

As one of their main contributions, \cite{DerakhshanGMM22} identified the optimal search policy for a single customer, with respect to any given assignment $\mathcal{A}: [n] \rightarrow [n]$. This policy begins by computing a monotonically-decreasing sequence of reservation prices $r_1 \geq \cdots \geq r_n$, where each price $r_p$ is the unique solution to $\expar{\ln(1+r_p+W)} - \ln(1+r_p) = c_p$, implying in particular that it is independent of the assignment ${\cal A}$.
Intuitively speaking, this price can be viewed as the deterministic reward that would make the customer indifferent between inspecting the product placed in position $p$ and stopping her search. With respect to these prices, the optimal policy dictates that the search procedure terminates at the first position $s^{\cal A}$ whose reservation price $r_{s^{\cal A}}$ is exceeded by the total preference weight of the products $\mathcal{A}(1), \ldots, \mathcal{A}(s^{\cal A})$. As explained in the next section, given these insights on customer behavior, the authors turn their attention to the retailer side, studying the algorithmic question of how products should be ranked when aiming to optimize either market share or consumer welfare.

\subsection{Formal model description} \label{subsec:model_description}
In what follows, we provide a complete mathematical description of the market share ranking problem, noting that the precise way by which one begins with the above-mentioned search policy and lands in the computational questions below can be better appreciated by consulting the work of \citet[Sec.~3-4]{DerakhshanGMM22}. For ease of presentation, the finer details of this setting will be conveyed in an incremental way, starting with its input parameters, moving on to explaining what our solution concept is and how customer consideration sets are formed, and ending with the  objective function to be optimized. Readers who are unfamiliar with MNL-based assortment optimization may benefit from the additional background on this topic in Section~\ref{related_works}.

\begin{figure}[ht]
    \centering
    \begin{subfigure}[b]{0.47\textwidth}
        \centering
        \includegraphics[width=\textwidth]{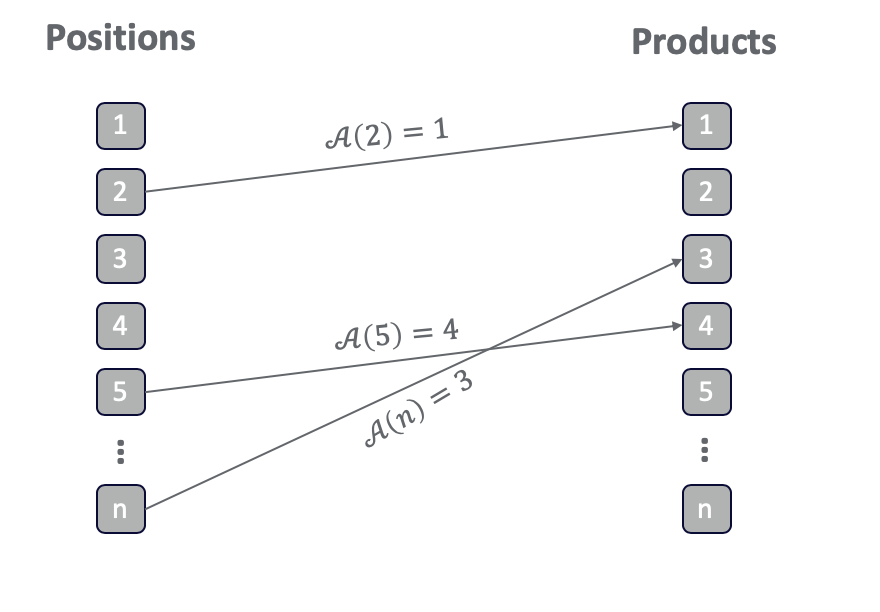}
        \caption{}
        \label{fig:figure1}
    \end{subfigure}
    \hfill
    \begin{subfigure}[b]{0.52\textwidth}
        \centering
        \includegraphics[width=\textwidth]{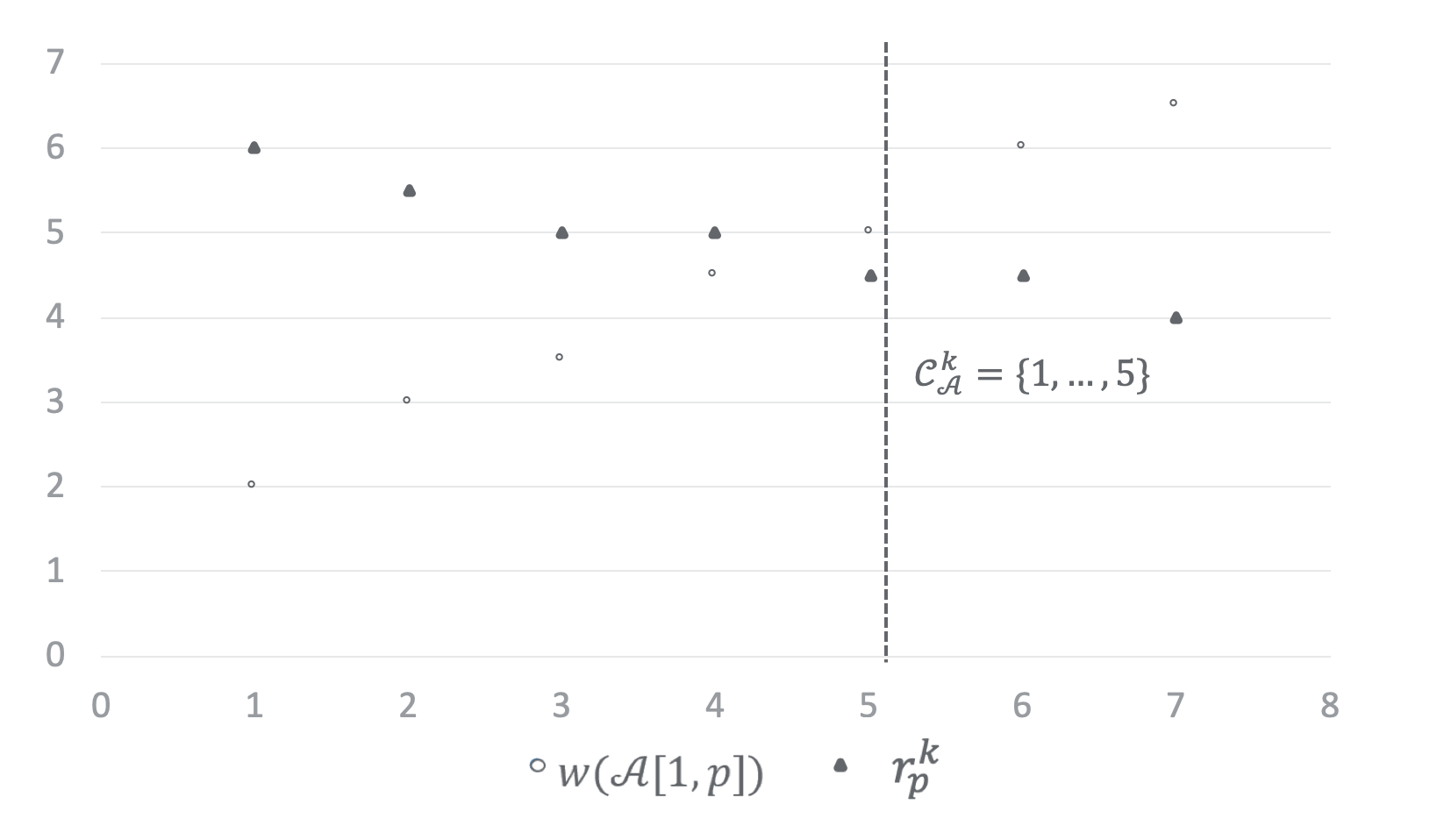}
        \caption{}
        \label{fig:figure2}
    \end{subfigure}
    \caption{(a) Illustration of the assignment $\mathcal{A}$; (b) The formation of a consideration set for a customer segment $k$.}
    \label{fig:mainfigure}
\end{figure}

\paragraph{Products and assignments.} In the market share ranking problem, we are given a finite set of products, designated by $1, \ldots, n$. We assume that each product $i \in [n]$ is associated with an MNL-based preference weight of $w_i$, whose precise role will be explained in the sequel. At a high level, our solution concept will be a complete linear order over these products. To this end, as illustrated in Figure~\ref{fig:figure1}, it is convenient to take the view of deciding how the underlying products should be spread across a sequence of so-called positions, $1, \ldots, n$.  At least intuitively, one should imagine that these positions dictate the level of visibility received by each product, with low-index positions being more accessible than high-index ones, similarly to how search query results are typically displayed by online platforms. To formalize this notion, our solution space will consist of position-to-product assignments, ${\cal A} : [n] \to [n]$. The latter function is bijective, defining a one-to-one correspondence in which ${\cal A}(p)$ stands for the unique product we decide to place at position $p$. 

\paragraph{Customer segments and their consideration sets.} From a demand perspective, we are provided with a known collection of customer segments, $1, \ldots, K$, whose respective proportions out of the entire population are denoted by $\theta_1, \ldots, \theta_K$. For notational brevity, we refer to each segment $k \in [K]$ simply as ``customer $k$''. With respect to any given assignment ${\cal A} : [n] \to [n]$, each of these customers makes a purchasing decision out of the offered products in two sequential steps:
\begin{enumerate}
    \item {\em Formation of consideration set}: Initially, the products on display are screened based on their positions. Specifically, each customer $k$ is associated with an individual position-dependent sequence of reservation prices, $r_1^k \geq \cdots \geq r_n^k$. Next, this customer constructs her consideration set $C_k^{\cal A}$ by inspecting the positions $1, \ldots, n$ one after the other, including each viewed product up to the point whose cumulative preference weight exceeds the reservation price of that position, as demonstrated in Figure~\ref{fig:figure2}. Formally, let us define the stopping point $s_k^{\cal A}$ of customer $k$ as the minimal index $p$ for which $w( {\cal A}[1,p] ) \geq r_p^k$, with the convention that $s_k^{\cal A} = n$ when $w( {\cal A}[1,n] ) < r_n^k$. Then, her consideration set is given by $C_k^{\cal A} = {\cal A}[1,s_k^{\cal A}]$, which is precisely the set of products placed by the assignment ${\cal A}$ up to and including the stopping point $s_k^{\cal A}$.  

    \item {\em MNL-driven choice}: Subsequently, customer $k$ either picks a single product out of her consideration set $C_k^{\cal A}$, or decides to avoid making any purchase; this choice is assumed to be governed by the Multinomial Logit model. In other words, each of the products $i \in C_k^{\cal A}$ is the one chosen with probability $\pi( i, C_k^{\cal A} ) = \frac{ w_i }{ 1 + w( C_k^{\cal A} )}$. For completeness, the residual probability is allocated to the no-purchase option, designated as ``product $0$'', meaning that $\pi( 0, C_k^{\cal A} ) = \frac{ 1 }{ 1 + w( C_k^{\cal A} )}$. 
\end{enumerate}
In light of this choice mechanism, it follows that the market share $M_k( {\cal A} )$ we capture with respect to customer $k$, standing for the probability of purchasing any of the offered products, is given by
    \[ M_k( {\cal A} ) ~~=~~ \frac{ w( C_k^{\cal A} ) }{ 1 + w( C_k^{\cal A} )} \ . \]

\paragraph{Objective function.} With all necessary ingredients in place, the market share ranking problem asks us to determine a position-to-product assignment ${\cal A} : [n] \to [n]$ whose expected market share $M( {\cal A} )$ is maximized. This expectation is taken with respect to the segment proportions $\theta_1, \ldots, \theta_K$, implying that our objective function can be written as
\[ M( {\cal A} ) ~~=~~ \sum_{k \in [K]} \theta_k \cdot M_k( {\cal A} ) \ . \]

\subsection{Existing results and open questions} \label{sec:prev_results_and_questions} As previously mentioned, a particularly insightful finding of \cite{DerakhshanGMM22} resides in identifying the optimal search policy for a single customer who wishes to maximize his expected welfare, resulting in the reservation-price-guided formation of consideration sets described in Section~\ref{subsec:model_description}. From an algorithmic perspective, this characterization gives rise to two basic optimization problems, where we wish to efficiently compute a product ranking that maximizes either market share or customer welfare. Since our work focuses on computational questions regarding the market share ranking problem, we proceed by reviewing the primary results of \citet{DerakhshanGMM22} in this context, with an emphasis on highlighting fundamental open questions. 

\paragraph{Intractability results.} In terms of lower bounds, \citet[Sec.~5]{DerakhshanGMM22}, established that the market share ranking problem is $\mathrm{NP}$-complete, meaning that an optimal ranking cannot be computed in polynomial time unless $\mathrm{P}=\mathrm{NP}$. At a high level, their proof maps instances of the $c$-bounded uniform knapsack problem to market share ranking instances with only two customer segments (i.e., $K=2$). It is worth mentioning that when $K=O(1)$, an FPTAS can  be designed by employing efficient enumeration methods, similarly to those presented in Section~\ref{sec:compute_partition}. The conjunction of these results leads to the natural question of whether dealing with an arbitrary number of customer segments is provably harder, specifically asking:

\begin{center}
{\em Is the market share ranking problem strongly NP-hard?}
\end{center}
An affirmative answer to this question would rule out the possibility of obtaining an FPTAS for general problem instances.  

\paragraph{Constant-factor approximations.} On the positive side, \cite{DerakhshanGMM22} proposed the ``$w$-ordering'' heuristic, where one ranks the underlying products by weakly-decreasing  preference weights. This algorithm was shown to produce a ranking whose expected market share is at least $\max \{\frac{M(\mathcal{A}^*)}{2}, M(\mathcal{A}^*)-0.1716\}$, with $\mathcal{A}^*$ standing for an optimal assignment. Such a heuristic is particularly appealing from an implementation perspective, since it is extremely simple and does not require any information about the customers' reservation prices. Knowing that the market share ranking problem admits a constant-factor approximation, our next question is:
\begin{quote}
{\em Can we obtain improved performance guarantees, perhaps by infusing new algorithmic ideas and analytic approaches?}
\end{quote}

\paragraph{Approximation scheme.} Utilizing a dynamic programming approach, \cite{DerakhshanGMM22} proceeded to provide an approximation scheme for the market share ranking problem. 
Their algorithm consists of two main steps: Bucketing and allocation. In the bucketing step, product weights are partitioned via a geometric grid, with each product placed in a bucket where its weight resides. In the allocation step, one determines a position-to-product assignment that maximizes market share, subject to operating under the low-weight priority (LWP) rule; here, among products in the same bucket, those with smaller weights are forced to appear in lower-indexed positions. Interestingly, this constraint was shown to be satisfied by at least one $(1-\eps)$-approximate assignment, and its structural simplicity allowed the authors to determine such an assignment by means of dynamic programming.

In terms of efficiency, this approach computes an assignment whose expected market share is within factor $1-\eps$ of optimal; however, its running time is $O(K^{2}n^{O( \frac{ 1 }{ \eps } \log \frac{w_{\max}}{w_{\min}} ) })$, where $w_{\max}$ and $w_{\min}$ respectively stand for the maximal and minimal preference weight of any product. Unfortunately, in arbitrarily-structured instances, $\frac{w_{\max}}{w_{\min}}$ may not be bounded by a constant, or even by a polynomial function of the input size, meaning that this algorithm corresponds to a polynomial-time approximation scheme (PTAS) only when $\frac{w_{\max}}{w_{\min}}=O(1)$. This limiting dependency leads to our next question:
\begin{quote}
{\em Is there a black-box reduction from general problem instances to those where $\frac{w_{\max}}{w_{\min}} = \mathrm{poly}(n, \frac{1}{\eps})$, losing only an $O(\eps)$-term in optimality?}
\end{quote}
Assuming an affirmative answer, the approximation scheme proposed by \cite{DerakhshanGMM22} would automatically be converted into a quasi-PTAS. Along these lines, the bucketing step of their approach, which unavoidably creates $\Omega(\frac{1}{\eps}\log \frac{w_{\max}}{w_{\min}})$ weight classes, directly translates to the dimension of the LWP-based dynamic program. This dependency represents an intrinsic bottleneck, steering us away from further exploring dynamic programming approaches where an exponent of $\Omega(\frac{1}{\eps}\log \frac{w_{\max}}{w_{\min}})$ appears to be inherent. Given these limitations, our final and most challenging questions are:
\begin{quote}
{\em Can we obtain a true PTAS for the market share ranking problem? What algorithmic methods will be useful in this context?}
\end{quote} 

\subsection{Main contributions}
The main contribution of this paper consists in devising a polynomial-time approximation scheme for the market share ranking problem, utilizing fresh technical developments and analytical ideas, in conjunction with revising some of the original insights fleshed out by \cite{DerakhshanGMM22}. Along the way, we introduce a black-box reduction, mapping general instances of the market share ranking problem into ``bounded ratio" instances, where $\frac{w_{\max}}{w_{\min}} = \mathrm{poly}(n, \frac{1}{\eps})$, showing that this result directly leads to an elegant and easily-implementable quasi-PTAS. Finally, to provide a complete computational characterization, we prove that the market share ranking problem is strongly $\mathrm{NP}$-hard. In what follows, we elaborate on the specifics of these contributions in greater detail. 

\paragraph{Hardness results.} 
In Section~\ref{sec:complement}, we prove that the market share ranking problem is strongly NP-hard, thereby excluding the possibility of obtaining an FPTAS. We mention in passing that, as opposed to the NP-hardness proof of \cite{DerakhshanGMM22}, our reduction produces market share ranking instances with an arbitrary number of customer segments.  

\begin{restatable}{restatable_thm}{hardnessthm}\label{thm_hardness}
The market share ranking problem is strongly NP-hard.
\end{restatable}

This result is derived via a carefully-crafted reduction from the restricted-input 3-partition problem, where briefly put, one wishes to determine whether a set of integers $a_1, \ldots a_{3K} \in (\frac{T}{4}, \frac{T}{2})$ can be partitioned into $K$ triplets, each summing up to $T$. At least intuitively, the main theme behind our proof consists in deploying $a_1, \ldots, a_{3K}$ as the preference weights of $3K$ products. To bridge between the two problems, we manipulate the stopping criterion through which consideration sets are formed. Specifically, by defining an appropriate sequence of reservation prices, along with additional virtual products, we force each customer $k \in [K]$ to stop right at the $k$-th triplet, thereby creating an analogy between customers and triplets. 

\paragraph{Warm up: Approximation scheme in quasi-polynomial time.}
In Section~\ref{subsec:qptas}, we devise a black-box reduction, mapping general instances of the market share ranking problem into what we call ``bounded-ratio" instances,  where $\frac{w_{\max}}{w_{\min}} = \mathrm{poly}(n, \frac{1}{\eps})$. The guiding principle behind this result is that products with very small preference weights can be shown to have a negligible effect on the overall market share. Therefore, our reduction uniformly treats such products by rounding their preference weights up to a threshold that depends on $w_{\max}, n$, and $\frac{1}{\eps}$, making the updated ratio  $\frac{w_{\max}}{w_{\min}}$ polynomial in $n$ and $\frac{1}{\eps}$. In turn, we proceed to utilize the approximation scheme of \cite{DerakhshanGMM22}, ending up with a near-optimal assignment in $O ( n^{ O_{\eps}( \log n ) } )$ time.

\begin{restatable}{restatable_thm}{thmqptas}\label{thm_qptas}
For any $\eps > 0$, the market share ranking problem can be approximated within factor $1 - \eps$ of optimal. The running time of our algorithm is $O ( K^2 n^{ O(\frac{1}{\eps}\log\frac{n}{\eps}) } )$.
\end{restatable}

\paragraph{Main result: Approximation scheme in truly polynomial time.} 
In Sections~\ref{sec:overview} and~\ref{sec:compute_partition}, we devise a polynomial-time approximation scheme (PTAS) for the market share ranking problem, which is the best possible approximability result one can hope for, given Theorem~\ref{thm_hardness}. As a side note, for simplicity of presentation, we have not made any effort at optimizing $\frac{1}{\eps}$-dependencies in the running time exponent mentioned below.

\begin{restatable}{restatable_thm}{maintheorem}
\label{thm:main_result}
For any $\eps > 0$, the market share ranking problem can be approximated within factor $1 - \eps$ of optimal. The running time of our algorithm is $O(n^{O(\frac{1} {\eps^{6}}\log\frac{1}{\eps})})$.
\end{restatable}

Our approach introduces several novel ingredients on which we proceed to succinctly elaborate next. First, similarly to \cite{DerakhshanGMM22}, we examine the family of ``sorted-within-class'' assignments, which is known to contain at least one near-optimal assignment. However, we further explore and exploit their  structural properties to decompose any such assignment into ``blocks'', where contiguous sequences of products are grouped based on cumulative weight thresholds. In turn, our key algorithmic innovation resides in shifting focus from directly computing position-to-product assignments to determining a highly-structured partition of the underlying products, which can eventually be mapped into a concrete assignment with a negligible loss in optimality. Without delving into technicalities, each subset in this partition adheres to appropriately-defined weight, cardinality, and sequencing constraints, closely mirroring the properties of its corresponding block. Once a suitable partition is computed, we provide an efficient method to convert this partition into an approximate sorted-within-class assignment. In terms of analysis, one particularly novel feature of our approach lies in creating a customer trichotomy according to their stopping points with respect to an optimal assignment, as dictated by the optimal search policy. For each of the three resulting groups, we propose a distinct mechanism to argue about the contribution of its customers toward the overall market share, allowing us to evaluate this contribution against the analogous quantity with respect to the optimal assignment.

\subsection{Related work}\label{related_works}
 Given that our work addresses computational questions at the intersection of logit-based models, display optimization, and incremental packing, we proceed by providing a succinct overview of key studies in these domains. The upcoming selection of papers is by no means intended to serve as an exhaustive literature review; rather, these papers offer a broader understanding of the intricacies involved in our problem of interest.

\paragraph{Assortment optimization under MNL preferences.} The Multinomial Logit (MNL) choice model has been one of the most extensively studied frameworks in revenue management, largely due to its computationally efficient parameter estimation methods (\cite{McFadden1974, HausmanMcFadden1984}), even when working with limited data \citep{Ford1957, Negahban2012}, and due to its computational tractability across various applications. Probably the most well-known contribution in this domain is that of \cite{TalluriVanRyzin04}, who devised a polynomial-time exact algorithm for the unconstrained assortment optimization problem. Subsequently, researchers proceeded to broaden the scope of this model by studying various structural constraints \citep{Rusmevichientong2010,Sumida2020, DBLP:journals/ior/DesirGZ22}, employing random choice parameters (\cite{Rusmevichientong2014}), investigating the notion of consideration sets (\cite{FeldmanTopaloglu}), and considering many additional directions. Building on these foundations, further research in assortment optimization has pursued models where time-dependent features are introduced to the framework \citep{Mahajan2001, FloresBerbegliaVanHentenryck2019, SegevFeldman22}. Additionally, robust optimization techniques have been explored as means of  deriving approximate solutions \citep{Rusmevichientong2012}.

\paragraph{Display optimization.} Display optimization problems have recently been emerging as a hot research topic in revenue management. In such scenarios, our objective is to determine an optimal display of items, given  heterogeneous consumers whose purchasing decisions are accordingly affected. Once items are assigned to their positions, customers are typically assumed to consider only a subset of these items, displayed in the most favorable positions, before picking an alternative through a given choice model. Among others, selected papers in this context include those of \cite{JamesWilliamson}, \cite{Gallego20}, and \cite{AouadS21}. Additional studies combine online settings with learning \citep{golrezaei2020learning, CaoSun2023}, and unveil hidden connections to submodular maximization \citep{asadpour2022sequential}. Finally, readers may be interested to learn about display optimization under various choice models and about some of their real-world applications by consulting the work of \cite{BreugelmansCampoGijsbrechts2007}, \cite{AgarwalHosanagarSmith2011}, \cite{Gallego20}, and \cite{AouadSaban2022}. 

\paragraph{Incremental packing.} Yet another topic in discrete optimization that bears certain lines of similarity with our work is incremental packing. Such settings can be viewed as multi-period extensions of classical packing and resource allocation problems, such as binary/integer knapsack, subset sum, geometrical packing, and many more. These extensions are meant to better capture real-life scenarios, such as sequentially choosing what resources should be allocated to a project over time, rather than making a single decision for the entire planning horizon. Under such a framework, we are typically facing dynamic capacity constraints that evolve over time, with the option of selecting additional items as time progresses. Inspiring some of our algorithmic methods is the work of \cite{AouadS23a} and \cite{FaenzaSZ23} on the incremental knapsack problem. Specifically, their analysis leverages efficient enumeration techniques that motivate the design of our polynomial-time approximation scheme. Additional papers in this domain study the inclusion of further structural assumptions, either attaining polynomial-time approximation schemes or employing LP-based approaches to obtain constant-factor approximations \citep{Bienstock2013, Adjiashvili2014, faenza2018ptas, DellaCroce2019}.

%% file: TEX-Hardness.tex
\section{Hardness Results} \label{sec:complement}

The main contribution of this section resides in pinpointing the precise sources of intractability hiding within market share ranking, arguing that this problem is strongly NP-hard. This result complements our algorithmic advances, showing that a polynomial-time approximation scheme is the best outcome one can hope for. Along these lines, Section~\ref{subsec:strong_np_hard} creates a bridge between market share ranking and the infamous 3-partition problem. Subsequently, Section~\ref{subsec:hardness_analysis} analyzes the computational consequences of this relationship, thereby migrating well-known strong NP-hardness results from the latter setting to the former. To streamline these ideas, several technical claims will be outsourced along the way to Sections~\ref{app:proof_lem_large_pos_large}-\ref{app:proof_clm_hardness_bound_other}.

\subsection{The reduction} \label{subsec:strong_np_hard}

In what follows, we describe a polynomial-time reduction from the $3$-partition problem to market share ranking. In particular, we will be exploiting the so-called restricted-input formulation of the former setting. While these two problems may appear to be unrelated, they actually share a number of well-hidden features that allow for this mapping. As explained below, the 3-partition problem involves deciding whether a given set of integers can be divided into triplets, all summing to the exact same quantity. Intuitively, the core idea of our proof lies in utilizing these integers as the preference weights of an appropriately-defined set of products, whereas customers will be representing potential triplets. To connect the two problems, we define a suitable sequence of reservation prices for each customer and introduce additional virtual products, ensuring that each customer halts her search at a unique triplet, with the best-possible expected market share telling us whether a $3$-partition exists or not.

\paragraph{Restricted-input $\bs{3}$-partition.}  Bringing readers up to speed, in this derivative of the 3-partition problem, we are given $3K$ positive integers, $a_1, \ldots, a_{3K}$, summing up to $KT$. Each of these integers is assumed to be residing within the open interval $(\frac{ T }{ 4 }, \frac{ T }{ 2 })$. Our objective is to decide whether there exists a partition of these integers into $K$ triplets $S_1, \ldots, S_K$, such that $a( S_1 ) = \cdots = a( S_K ) = T$. For additional background on this problem, one can consult the excellent book of \citet[Sec.~4.2]{GareyJ79}.

\paragraph{Reduction to market share.} Given an instance of this form, we construct a corresponding market share ranking instance as follows:
\begin{enumerate}
    \item {\em Products}: There are $4K$ products, of which those indexed by $1, \ldots, 3K$ will be referred to as ``small'', with the remaining $K$ being ``large''. Each small product $i \in [3K]$ has a preference weight of $w_i = a_i$, whereas all large products have a uniform weight of $L = 4(K^3T+1)$.

    \item {\em Customers and reservation prices}: We create $K$ customer segments, with the occurrence probability of each customer $k \in [K]$ being $\theta_k = \alpha \cdot \frac{ k(L+T)+1 }{ k(L+T) }$. Here, $\alpha$ is simply a normalization constant, meaning that $\alpha = (\sum_{k \in [K]} \frac{ k(L+T)+1 }{ k(L+T) })^{-1}$. In addition, this customer is associated with the following sequence of weakly-decreasing reservation prices:
    \begin{itemize}
        \item {\em Positions $1, \ldots, 4k-1$}: $r_1^k = \cdots = r_{4k-1}^k = (k-1)L + kT + \frac{1}{2}$.

        \item {\em Positions $4k, \ldots, 4K$}: $r_{4k}^k = \cdots = r_{4K}^k = 0$.
    \end{itemize}    
\end{enumerate}

\subsection{Analysis}\label{subsec:hardness_analysis}

Given the mapping described above, we proceed by relating between these two problems, showing that the existence of a 3-partition is equivalent to that of an assignment whose expected market share crosses a certain threshold. This result, whose specifics are given in the next claim, proves that the market share ranking problem is strongly NP-hard.

\begin{lemma}
There exists a $3$-partition $S_1, \ldots, S_K$ with $a( S_1 ) = \cdots = a( S_K ) = T$ if and only if there exists an assignment ${\cal A}$ with an expected market share of $M( {\cal A} ) \geq K \alpha$.
\end{lemma}

\paragraph{The easy direction: $\bs{3\text{-partition} \Rightarrow M( {\cal A} ) \geq K \alpha}$.} Suppose there exists a $3$-partition $S_1, \ldots, S_K$ with $a( S_1 ) = \cdots = a( S_K ) = T$. In this scenario, let us define an assignment ${\cal A}$ as follows:
\begin{itemize}
    \item The three small products corresponding to $S_1$ are placed at positions $1,2,3$ in arbitrary order. Similarly, skipping position $4$, those corresponding to $S_2$ are placed at positions $5,6,7$, so on and so forth.

    \item The $K$ large products appear in positions $4, 8, 12, \ldots, 4K$. Since all large products share an identical preference weight, their internal order within these positions is inconsequential.
\end{itemize}
With respect to this assignment, we observe that the stopping point $s^{\cal A}_k$ of each  customer $k \in [K]$ is precisely $4k$. This conclusion follows by noting that, up to position $4k-1$, we are placing $k-1$ large products as well as those corresponding to $\mathcal{S}_1, \ldots, \mathcal{S}_k$, and therefore 
\[ w( {\cal A}[1,4k-1] ) ~~=~~ (k-1)L + \sum_{\ell \in [k]} a(S_{\ell}) ~~=~~ (k-1)L + kT ~~=~~ r_{4k-1}^k - \frac{1}{2} \ , \]
implying that $s^{\cal A}_k \geq 4k$. On the other hand, since position $4k$ contains a large product,
\[ w( {\cal A}[1,4k] ) ~~=~~ w( {\cal A}[1,4k-1] )  + L~~=~~ k(L+T) ~~>~~ 0 ~~=~~ r_{4k}^k \ . \]
As a result, the consideration set of each customer $k$ turns out to be $C_k^{\cal A} = {\cal A}[1,4k]$, implying that our assignment has an expected market share of 
\[ M({\cal A}) ~~=~~ \sum_{k \in [K]} \theta_k \cdot M_k( {\cal A}) ~~=~~ \sum_{k \in [K]} \theta_k \cdot \frac{ k(L+T) }{ 1 + k(L+T) } ~~=~~ K \alpha \ , \]
where the last equality holds since $\theta_k = \alpha \cdot \frac{ k(L+T)+1 }{ k(L+T) }$.

\paragraph{The difficult  direction: $\bs{M( {\cal A} ) \geq K \alpha \Rightarrow 3\text{-partition}}$.} Let ${\cal A}^*$ be an optimal assignment, and suppose that $M( {\cal A}^* ) \geq K \alpha$. With respect to this assignment, let $p_1^* < \cdots < p_K^*$ be the sequence of positions to which ${\cal A}^*$ assigns large products. The next claim, whose proof is provided in Section~\ref{app:proof_lem_large_pos_large}, shows that the very last position, $4K$, is necessarily holding a large product. At a high level, to establish this result, we argue that relocating the latest large product to position $4K$ may only expand the consideration set of customer $K$, while leaving those of customers $1, \ldots, K-1$ unchanged, thereby contradicting the optimality of $\mathcal{A}^*$.

\begin{lemma} \label{lem:large_pos_large}
$p_K^* = 4K$.
\end{lemma}

Now, let $W_1^*$ be the total weight of the products placed by ${\cal A}^*$ before the first large product, i.e., $W_1^* = w( {\cal A}^*[1,p_1^*-1] )$. Similarly, let $W_2^*$ be the total weight of those placed between the first and second large products, namely, $W_2^* = w( {\cal A}^*[p_1^*+1, p_2^*-1] )$. We define $W_3^*, \ldots, W_K^*$ in a similar way. By Lemma~\ref{lem:large_pos_large}, we know that the optimal assignment ${\cal A}^*$ does not place any product after the $K$-th large product, meaning that $\sum_{k \in [K]} W_k^* = KT$. The crux of our reduction is summarized in Lemma~\ref{lem:harness_buckets_equal} below, showing that $W^*_1, \ldots, W^*_K$ must be equal. To derive this result, we first show that, for all $k \in [K]$, the cumulative weight $\sum_{\kappa \leq k} W_{\kappa}^*$ cannot exceed $kT$; otherwise, we establish a contradiction to $M(\mathcal{A}^*) \geq K\alpha$. Subsequently, we prove that $\sum_{\kappa \leq k} W_{\kappa}^* \leq kT$ always holds with equality, as any strict inequality for a specific $k \in [K]$ would again lead to a contradiction of $M(\mathcal{A}^*) \geq K\alpha$. 

\begin{lemma} \label{lem:harness_buckets_equal}
$W_1^* = \cdots = W_K^* = T$.    
\end{lemma}

Consequently, letting $S_1 = \{ a_i : i \in {\cal A}^*[1,p_1^*-1] \}, S_2 = \{ a_i : i \in {\cal A}^*[p_1^*+1, p_2^*-1] \}, \ldots, S_K = \{ a_i : i \in {\cal A}^*[p_{K-1}^*+1, p_K^*-1] \}$, we have just defined a partition of $a_1, \ldots, a_{3K}$ satisfying $a( S_1 ) = \cdots = a( S_K ) = T$. Moreover, since each of these integers resides in $(\frac{ T }{ 4 }, \frac{ T }{ 2 })$, the subsets $S_1, \ldots, S_K$ are necessarily triplets.

\subsection{Proof of Lemma~\ref{lem:large_pos_large}} \label{app:proof_lem_large_pos_large}

Our proof is based on arguing that, when $p^*_K < 4K$, the assignment $\mathcal{A}^*$ cannot be optimal. To this end, let us define a modified assignment, ${\cal A}$, where the $K$-th large product in order of appearance is relocated to position $4K$, pulling back the products ${\cal A}^*(p_K^* + 1), \ldots, {\cal A}^*(4K)$ by a single position. In other words,
\[ {\cal A}(p) ~~=~~
\begin{dcases}
{\cal A}^*(p), & \text{if } p \leq p_K^* - 1 \\
{\cal A}^*(p+1) , \quad & \text{if } p_K^* \leq p \leq 4K - 1 \\
{\cal A}^*(p_K^*), & \text{if } p = 4K
\end{dcases} \]
In the remainder of this proof, we show that $M( {\cal A} ) > M( {\cal A}^* )$, contradicting the optimality of ${\cal A}^*$. For this purpose, we argue that the consideration sets of customers $1, \ldots, K-1$ remain unchanged, namely, $C_k^{\cal A} = C_k^{{\cal A}^*}$ for every $k \in [K-1]$. In addition, we show that $C_K^{{\cal A}^*}$ is a proper subset of $C_K^{\cal A}$. Since all product weights and segment probabilities are strictly positive, these claims jointly imply that $M( {\cal A} ) > M( {\cal A}^* )$, as desired.

First, to prove that $C_k^{\cal A} = C_k^{{\cal A}^*}$ for every $k \in [K-1]$, it suffices to show that $s_k^{{\cal A}^*} \leq p_k^*$, since the assignment $\mathcal{A}$ keeps positions $1, \ldots, p^*_{K}-1$ unchanged in comparison to $\mathcal{A}^*$. This inequality is obtained by noting that 
\begin{eqnarray}\label{eqn_sk<pk}
        w( {\cal A}^*[1,p_k^*]) &\geq& kL \label{sk<pk:step1}\\
        &\geq& (k-1)L + kT + \frac{1}{2} \label{sk<pk:step2}\\
        &\geq& r^k_{ p_k^* }  \label{sk<pk:step3}\ . \
\end{eqnarray}
Here, inequality~\eqref{sk<pk:step1} holds since ${\cal A}^*[1,p_k^*]$ contains at least $k$ large products. Inequality~\eqref{sk<pk:step2} follows from the observation that $L \geq kT +\frac{1}{2}$, since $L = 4(K^3T+1)$. Finally, inequality~\eqref{sk<pk:step3} is obtained by noting that $r^k_{p} \leq (k-1)L + kT + \frac{1}{2}$, for every $p \in [4K]$.

Second, to show that $C_K^{\cal A} \supsetneq C_K^{{\cal A}^*}$, it suffices to argue that $s_K^{{\cal A}} > s_K^{{\cal A}^*}$. To this end, note that $w( {\cal A}^*[1,p_K^*]) \geq r^K_{ p_K^* }$, since inequality~\eqref{sk<pk:step3} is applicable to $k = K$ as well. Consequently, $s_K^{{\cal A}^*} \leq p_K^* < 4K$, by our initial assumption regarding $p^*_K$. On the other hand, $s^{\cal A}_K = 4K$, since within positions $1, \ldots, 4K-1$, the assignment $\mathcal{A}$ has products $1, \ldots, 3K$ as well as $K-1$ large products, and therefore,
\[ w( {\cal A}[1,4K-1]) ~~=~~ (K-1)L + KT ~~=~~ r^K_{ 4K-1 }-\frac{1}{2} \ . \]

\subsection{Proof of Lemma~\ref{lem:harness_buckets_equal}} \label{app:proof_lem_harness_buckets_equal}

To establish the desired claim, we will first argue that $\sum_{\kappa \leq k} W_{\kappa}^* \leq kT$, for every $k \in [K]$. Given this result, our second step will consist of showing that none of these inequalities can be strict, implying as an immediate corollary that $W_1^* = \cdots = W_K^* = T$.

\paragraph{Step 1: $\bs{\sum_{\kappa \leq k} W_{\kappa}^* \leq kT}$, for every $\bs{k \in [K]}$.} Suppose on the contrary that there exists some $\ell \in [K]$ for which $\sum_{\kappa \leq \ell} W_{\kappa}^* > \ell T$. Our analysis is based on the next two claims, whose respective proofs are deferred to Sections~\ref{app:proof_clm_hardness_bound_ell} and~\ref{app:proof_clm_hardness_bound_other}.

\begin{claim} \label{clm:hardness_bound_ell}
$\theta_{ \ell } \cdot M_{ \ell }( {\cal A}^* ) \leq \alpha \cdot ( 1 - \frac{ 1 }{ 2K^2 L} )$.
\end{claim}

\begin{claim} \label{clm:hardness_bound_other}
$\theta_k \cdot M_k( {\cal A}^* ) \leq~ \alpha \cdot ( 1 + \frac{ 2T }{ L^2 } )$, for every $k \neq \ell$.
\end{claim}

With these bounds in place, we are ready to reveal the resulting contradiction, arguing that $M( {\cal A}^* ) < K \alpha$. The latter inequality follows by noting that
\begin{eqnarray}
M( {\cal A}^* ) & = & \sum_{k \neq \ell} \theta_k \cdot M_k( {\cal A}^* ) + \theta_{ \ell } \cdot M_{ \ell }( {\cal A}^* ) \nonumber \\
& \leq & (K-1) \cdot \alpha \cdot \left( 1 + \frac{ 2T }{ L^2 } \right) + \alpha \cdot \left( 1 -  \frac{ 1 }{ 2K^2L } \right) \label{eq:step1} \\
& \leq & K \alpha + \alpha \cdot \left( \frac{ 2KT }{ L^2 } -  \frac{ 1 }{ 2K^2L }\right) \nonumber \\
& = & K \alpha + \alpha \cdot   \frac{ 4K^3T -L }{ 2K^2L^2 } \nonumber \\
& < & K \alpha \ . \label{eq:final}
\end{eqnarray}
Here, inequality~\eqref{eq:step1} is obtained by plugging in Claims~\ref{clm:hardness_bound_ell} and~\ref{clm:hardness_bound_other}, whereas  inequality~\eqref{eq:final} holds since $L = 4(K^3T+1)$.

\paragraph{Step 2: $\bs{\sum_{\kappa \leq k} W_{\kappa}^* = kT}$, for every $\bs{k \in [K]}$.} We begin by recalling that, as explained within the proof of Lemma~\ref{lem:large_pos_large}, the stopping point $s_k^{{\cal A}^*}$ of each customer $k \in [K]$ with respect to the optimal assignment ${\cal A}^*$ occurs no later than position $p_k^*$. Consequently, for the consideration set of this customer, we have $w( C_k^{{\cal A}^*} ) \leq kL + \sum_{\kappa \leq k} W_{\kappa}^* \leq k(L+T)$, where the last inequality follows from step~1. Now suppose that $\sum_{\kappa \leq \ell} W_{\kappa}^* < \ell T$, for some $\ell \in [K]$, meaning that $w( C_{\ell}^{{\cal A}^*} ) < \ell(L+T)$. These two inequalities bring us once again to a contradiction of the form $M( {\cal A}^* ) < K \alpha$, since 
\begin{eqnarray*}
M( {\cal A}^* ) & = & \sum_{k \neq \ell} \theta_k \cdot M_k( {\cal A}^* ) + \theta_{ \ell } \cdot M_{ \ell }( {\cal A}^* ) \\
& < & \sum_{k \neq \ell} \theta_k \cdot \frac{ k(L+T) }{ 1 + k(L+T) } + \theta_{\ell} \cdot \frac{ \ell(L+T) }{ 1 + \ell(L+T) } \\
& = & K \alpha \ , \ 
\end{eqnarray*}
where the last inequality holds since $\theta_k = \alpha \cdot \frac{ k(L+T)+1 }{ k(L+T) }$ for every $k \in [K]$.

\subsection{Proof of Claim~\ref{clm:hardness_bound_ell}} \label{app:proof_clm_hardness_bound_ell}

Our proof begins by examining the relation between the stopping point $s_{\ell}^{ {\cal A}^* }$ of customer $\ell$ and the position $p_{\ell}^*$, where the $\ell$-th large product resides. Specifically, we argue that $s_{\ell}^{ {\cal A}^* } \leq p_{\ell}^* - 1$, since
\begin{eqnarray}
w( {\cal A}^*[1, p_{\ell}^*-1] ) & = & w( {\cal A}^*[1, p_{\ell}^*] ) - L \nonumber \\
& = & \ell L + \sum_{\kappa \leq \ell} W_{\kappa}^* - L \nonumber \\
& \geq & (\ell-1)L + \ell T + 1 \label{eq:step2} \\
& > & r^{\ell}_{ p_{\ell}^*-1 }  \label{eq:step3}\ . 
\end{eqnarray}
Here, inequality~\eqref{eq:step2} follows by recalling that $\sum_{\kappa \leq \ell} W_{\kappa}^* > \ell T$ and that $\ell T, W_1^*, \ldots, W_K^*$ are all integers. Inequality~\eqref{eq:step3} holds since $r^{\ell}_p \leq (\ell-1)L+\ell T +\frac{1}{2}$ for every $p \in [4K]$.

As a result, we conclude that the consideration set $C_{\ell}^{ {\cal A}^* }$ of customer $\ell$ contains at most $\ell-1$ large products, implying that $w( C_{\ell}^{ {\cal A}^* } ) \leq (\ell-1)L + KT$. In turn,
\begin{eqnarray}
\theta_{ \ell } \cdot M_{ \ell }( {\cal A}^* ) & \leq & \alpha \cdot \frac{ \ell(L+T)+1 }{ \ell(L+T) } \cdot \frac{ (\ell-1)L + KT }{ 1 + (\ell-1)L + KT} \nonumber \\
& = & \alpha \cdot \left( 1 +  \frac{ 1 }{ \ell(L+T) } \right) \cdot \left( 1 - \frac{ 1 }{ 1 + (\ell-1)L + KT} \right) \nonumber \\
& \leq & \alpha \cdot \left( 1 +  \frac{ 1 }{ \ell L } \right) \cdot \left( 1 - \frac{ 1 }{ (\ell-1/2)L} \right) \label{eq:step4} \\
& \leq & \alpha \cdot \left( 1 +  \frac{ 1 }{ \ell L } - \frac{ 1 }{ (\ell-1/2)L} \right) \nonumber \\
& = & \alpha \cdot \left( 1 -  \frac{ 1 }{ 2\ell(\ell-1/2)L }  \right) \nonumber \\
& \leq & \alpha \cdot \left( 1 -  \frac{ 1 }{ 2K^2L }  \right) \ , \nonumber
\end{eqnarray}
where inequality~\eqref{eq:step4} can easily be verified by recalling that $L = 4(K^3T+1)$.

\subsection{Proof of Claim~\ref{clm:hardness_bound_other}} \label{app:proof_clm_hardness_bound_other}

We first observe that, for every customer $k \in [K]$,
\begin{eqnarray}
    w( C_k^{ {\cal A}^* } ) &=& w(\mathcal{A}^*[1,s^{\cal A^*}_k-1]) + w_{\mathcal{A}^*(s^{\cal A^*}_k)} \nonumber \\
    &\leq& r^k_{s^{\cal A^*}_k-1} + L \nonumber \\
    &\leq & k(L+T) +1 \ . \nonumber 
\end{eqnarray}
where the last inequality holds since $r^{ k }_{ p } \leq (k-1)L + kT + \frac{1}{2}$ for every position $p \in [4K]$. Therefore,
\begin{eqnarray*}
\theta_k \cdot M_k( {\cal A}^* ) & \leq & \alpha \cdot \frac{ k(L+T)+1 }{ k(L+T) } \cdot \frac{ kL + (k+1)T }{ 1 + kL + (k+1)T} \\
& = & \alpha \cdot \left( 1 +  \frac{ 1 }{ k(L+T) } \right) \cdot \left( 1 - \frac{ 1 }{ 1 + kL + (k+1)T} \right) \\
& \leq & \alpha \cdot \left( 1 +  \frac{ 1 }{ k(L+T) } \right) \cdot \left( 1 - \frac{ 1 }{ k(L+T) + 2T} \right) \\
& \leq & \alpha \cdot \left( 1 +  \frac{ 1 }{ k(L+T) } - \frac{ 1 }{ k(L+T) + 2T} \right) \\
& = & \alpha \cdot \left( 1 +  \frac{ 2T }{ k(L+T)(k(L+T) + 2T) } \right) \\
& \leq & \alpha \cdot \left( 1 +  \frac{ 2T }{ L^2 } \right) \ .
\end{eqnarray*}

%% file: TEX-QPTAS.tex
\section{Quasi-Polynomial-Time Approximation Scheme} \label{subsec:qptas}

In this section, we present a quasi-polynomial-time approximation scheme for the market share ranking problem, as formally stated in Theorem~\ref{thm_qptas}, repeated below. To this end, we devise an elegant reduction, mapping arbitrary market share ranking instances to so-called ``bounded-ratio'' instances, where the ratio between the extremal preference weights is polynomial in $n$ and $\frac{ 1 }{ \eps }$. Combined with the approximation scheme of \cite{DerakhshanGMM22}, which is exponential in $\log\frac{w_{\max}}{w_{\min}}$ by itself, we will argue that this reduction leads to a quasi-polynomial-time approximation scheme. 

\thmqptas*

\subsection{Algorithmic outline}\label{subsec:qptas_cstrct}
\paragraph{Putting aside the trivial case.} Let us first observe that, for any given instance $\mathcal{I}$ where the maximal preference weight is sufficiently large, specifically being $w_{\max} > \frac{ 1 }{ \eps }$, one can easily obtain a $(1-\eps)$-approximation. In particular, letting $i^*$ be the heaviest product, consider any assignment $\mathcal{A}$ that places $i^*$ at the most visible position, meaning that $\mathcal{A}(1)= i^*$. For such an assignment, each customer $k \in [K]$ is guaranteed to have this product within her consideration set $C^{\mathcal{I}, \mathcal{A}}_k$, possibly with additional products, leading to a market share of  
\begin{equation}
    M_{\mathcal{I}}(\mathcal{A}) ~~=~~ \sum_{k=1}^{K} \theta_k \cdot \frac{w(C^{\mathcal{I}, \mathcal{A}}_k)}{1 + w(C^{\mathcal{I},\mathcal{A}}_k)} ~~\geq~~ \frac{w_{\max}}{1+w_{\max}} \ . \label{obs_non_triv}
\end{equation}
We mention in passing that this lower bound on $M_{\cal I}(\mathcal{A})$ holds unconditionally, regardless of how $w_{\max}$ and $\frac{ 1 }{ \eps }$ are related. However, when $w_{\max} > \frac{ 1 }{ \eps }$, inequality~\eqref{obs_non_triv} further implies that 
\[ M_{\mathcal{I}}(\mathcal{A}) ~~\geq~~ \frac{1}{1+\epsilon} ~~\geq~~ (1-\epsilon)\cdot\opt(\mathcal{I}) \ . \] 
In the remainder of this section, we consider the non-trivial case, where $w_{\max} \leq \frac{ 1 }{ \eps }$.

\paragraph{Modifying product weights.} Given an arbitrarily structured instance ${\cal I}$, the reduction we propose operates by rounding up the preference weight of very low-attraction products.  Specifically, letting $\delta = \frac{\epsilon^2}{2n}$, our modified product weights $\{ \tilde{w}_i \}_{i \in [n]}$ are determined by setting
\[ \tilde{w_i} ~~=~~ \begin{dcases}
\delta w_{\max}, \quad & \text{if } w_i \leq \delta w_{\max} \\
w_i, \quad & \text{otherwise} 
\end{dcases} \]
In other words, all products of weight at most $\delta w_{\max}$ have their weight rounded up to this quantity. Beyond the latter alteration, our resulting instance $\tilde{\cal I}$ is identical to ${\cal I}$.

In what follows, we argue that the expected market share $\opt( \tilde{\mathcal{I}} )$ attained by an optimal assignment with respect to our modified instance does not lag much behind that of an optimal assignment for the original instance, $\opt( \mathcal{I} )$. At least intuitively, this relation is derived by arguing that, when very low-weight products are relocated to the ``end'' of an optimal assignment (i.e., high-index positions), each customer will still be keeping all high-weight products within her consideration set. Concurrently, the cumulative loss due to potentially dropping low-weight products will not significantly affect the expected market share of any customer. The proof of this result appears in Section~\ref{app:proof_qptas_opt_maintain}. 

\begin{lemma} \label{lem:qptas_opt_maintain}
$\opt( \tilde{\mathcal{I}} ) \geq (1-\epsilon) \cdot \opt( \mathcal{I} )$. 
\end{lemma}

\paragraph{Computing an approximate assignment $\bs{\tilde{\mathcal{A}}}$ for $\bs{\tilde{\cal I}}$.} As explained in Section~\ref{sec:prev_results_and_questions}, with respect to the modified instance $\tilde{\cal I}$, the approximation scheme of \cite{DerakhshanGMM22} can be implemented in $O(K^{2}n^{ O( \frac{1}{\eps}\log \tilde{\rho} ) })$ time, where $\tilde{\rho} = \frac{ \tilde{w}_{\max}}{\tilde{w}_{\min}}$ stands for the ratio between the extremal weights of this instance. However, by rounding product weights as described above, we have $\tilde{w}_{\min} \geq \delta {w}_{\max} = \frac{ \eps^2 }{ 2n } \cdot {\tilde{w}}_{\max}$, meaning in turn that $\tilde{\rho} \leq \frac{ 2n }{ \eps^2 }$. Consequently, within an overall running time of $O(K^{2}n^{ O( \frac{1}{\eps} \log \frac{n}{\eps} ) })$, their algorithm is guaranteed to compute an assignment $\tilde{\cal A}$ whose expected market share is $M_{ \tilde{\cal I} }( \tilde{\cal A} ) \geq (1 - \eps) \cdot \opt( \tilde{\cal I} )$. 

\paragraph{Why $\bs{\tilde{\cal A}}$ is near-optimal for $\bs{\cal I}$?} We conclude our analysis by showing that the assignment $\tilde{\mathcal{A}}$ is actually near-optimal with respect to the original instance ${\cal I}$. The intuition behind this result is that, since $\tilde{w}_i \geq w_i$ for every product $i \in [n]$, the stopping point of any customer with respect to the original instance ${\cal I}$ cannot occur later than her stopping point relative to the modified instance $\tilde{ \cal I}$. In other words, when we revert back to the original instance, the consideration set of each customer can only expand. Moreover, we will show that restoring the original product weights has a negligible impact on the overall market share. The formal proof of this claim is provided in Section~\ref{app:proof_lem_tildeA_for_I}.    

\begin{lemma} \label{lem:tildeA_for_I}
$M_{\cal I}( \tilde{\cal A} ) \geq (1 - 3 \eps) \cdot \opt( {\cal I} )$.
\end{lemma}

\subsection{Proof of Lemma~\ref{lem:qptas_opt_maintain}} \label{app:proof_qptas_opt_maintain}

Let ${\cal A}^*$ be an optimal assignment for the original instance $\mathcal{I}$. To establish the desired claim, we show that ${\cal A}^*$ can be converted into an assignment $\mathcal{A}$ for the modified instance $\tilde{\cal I}$ whose expected market share is $M_{ \tilde{\cal I} }( {\cal A} ) \geq (1 - \eps) \cdot M_{ {\cal I} }( {\cal A}^* )$. 

\paragraph{Constructing $\bs{\cal A}$.} In order to define ${\cal A}$, let us make use of $\mathcal{H} = \{ i \in [n] : w_i > \delta w_{\max} \}$ to designate the set of heavy products. With this notation, the assignment ${\cal A}$ is created by having all products in $\mathcal{H}$ remain in the exact same internal order as in ${\cal A}^*$, while arbitrarily organizing the set of light products ${\cal L} = [n] \setminus {\cal H}$ in the $n - |{\cal H} |$ highest-index positions. In other words, letting ${\cal H} = \{i_1, \ldots, i_{|{\cal H} |}\}$, with the convention that ${\cal A}^{*-1}( i_1 ) < \cdots < {\cal A}^{*-1}( i_{|{\cal H} |} )$, we set $\mathcal{A}(p) = i_p$ for every position $p \leq | {\cal H} |$. Then, each of the positions $|{\cal H}|+1, \ldots, n$ is assigned with a distinct product from ${\cal L}$ in arbitrary manner.

\paragraph{Properties of consideration sets.} Focusing on a single customer $k \in [K]$, we make use of $C^{\mathcal{I},\mathcal{A}^*}_k$ to designate her consideration set with respect to the optimal assignment ${\cal A}^*$ for ${\cal I}$. Similarly, $C^{\tilde{\mathcal{I}},\mathcal{A}}_k$ will denote her consideration set with respect to the assignment ${\cal A}$ for $\tilde{\cal I}$. We proceed by showing that all heavy products appearing in the former set also appear in the latter, namely, 
\begin{equation} \label{eqn:qptas_opt_maintain_eq0}
(C^{\mathcal{I}, \mathcal{A^*}}_k \cap \mathcal{H}) ~~\subseteq~~ (C^{\tilde{\mathcal{I}}, \mathcal{A}}_k \cap \mathcal{H}) \ .    
\end{equation}
To this end, letting $i_1, \ldots, i_h$ be the collection of products in $C^{\mathcal{I}, \mathcal{A}^*}_k \cap \mathcal{H}$, by recalling that ${\cal A}[1,h] = \{ i_1, \ldots, i_h \}$, it suffices to show that the stopping point $s^{\tilde{\cal I}, \mathcal{A}}_k$ of customer $k$ with respect to the assignment ${\cal A}$ occurs at position $h$ or later. This claim will follow by arguing that the stopping condition of customer $k$ is not met in any of the positions $1, \ldots, h-1$, which is equivalent to having $\tilde{w}( {\cal A}[1, h-1] ) < r^k_{h-1}$. To derive the latter inequality, note that
\begin{eqnarray}
        \tilde{w}( {\cal A}[1,h-1] ) \nonumber & = & \sum_{p \in [h-1]} w_{i_p}  \\ 
        & \leq & w( {\cal A}^*[1,s^{\mathcal{I}, {\cal A}^*}_k-1] ) \label{eqn:qptas_opt_maintain_eq1} \\ 
        & < & r^k_{ s^{\mathcal{I}, {\cal A}^*}_k-1 } \label{eqn:qptas_opt_maintain_eq2} \\
        & \leq & r^k_{h-1} \ . \label{eqn:qptas_opt_maintain_eq3}
\end{eqnarray}
Here, inequality~\eqref{eqn:qptas_opt_maintain_eq1} holds since $\{ i_1, \ldots, i_{h} \} \subseteq \mathcal{C}^{\mathcal{I}, \mathcal{A}^*}_{k} = {\cal A}^*[1,s^{\mathcal{I}, {\cal A}^*}_k]$. Inequality~\eqref{eqn:qptas_opt_maintain_eq2} is obtained by observing that, in the original instance, $s^{\mathcal{I}, {\cal A}^*}_k$ is the stopping point of customer $k$ with respect to ${\cal A}^*$, meaning that we still have $w( {\cal A}^*[1,s^{\mathcal{I}, {\cal A}^*}_k-1] ) < r^k_{ s^{\mathcal{I}, {\cal A}^*}_k-1 }$ at position $s^{\mathcal{I}, {\cal A}^*}_k-1$. Finally, inequality~\eqref{eqn:qptas_opt_maintain_eq3} follows by recalling that $r_1^k \geq \cdots \geq r_n^k$, and since we clearly have $s^{\mathcal{I}, {\cal A}^*}_k \geq h$.    

\paragraph{Relating between $\bs{M_{\tilde{\mathcal{I}}}(\mathcal{A})}$ and $\bs{M_{{\mathcal{I}}}(\mathcal{A}^*)}$.} In summary, to compare between the total weight of $C^{\tilde{\mathcal{I}}, \mathcal{A}}_k$ and $C^{{\mathcal{I}}, \mathcal{A}^*}_k$, based on relation~\eqref{eqn:qptas_opt_maintain_eq0}, we infer that 
\begin{eqnarray}
\tilde{w} \left( C^{\tilde{\mathcal{I}}, \mathcal{A}}_k \right) & \geq & \tilde{w} \left( C^{\tilde{\mathcal{I}}, \mathcal{A}}_k \cap \mathcal{H} \right) \nonumber \\
& \geq & \tilde{w} \left( C^{{\mathcal{I}}, \mathcal{A}^*}_k \cap \mathcal{H} \right) \nonumber \\
& \geq & \tilde{w} \left( C^{{\mathcal{I}}, \mathcal{A}^*}_k \right) - \tilde{w} ( {\cal L} )\nonumber \\
&\geq& w \left( C^{{\mathcal{I}}, \mathcal{A}^*}_k \right) - n \delta w_{\max} \ , \label{eqn:tildew_w_cons}
\end{eqnarray}
where the last inequality holds since $\tilde{w}_i \geq w_i$ for every product $i \in [n]$, and since $\tilde{w}_i = \delta w_{\max}$ for every $i \in {\cal L}$, by definition. 

Consequently, it follows that the expected market share of the assignment ${\cal A}$ with respect to the modified instance $\tilde{\cal I}$ is
\begin{eqnarray}
M_{\tilde{\mathcal{I}}}(\mathcal{A}) & = &  \sum_{k \in [K]} \theta_k \cdot \frac{ \tilde{w}(C_k^{\tilde{\mathcal{I}},\mathcal{A}})}{1+\tilde{w}(C_k^{\mathcal{\tilde{I}} ,\mathcal{A}})} \nonumber \\
& \geq & \sum_{k \in [K]} \theta_k \cdot \frac{ w(C_k^{\mathcal{I},\mathcal{A}^*})- n \delta w_{\max}}{1+w(C_k^{\mathcal{I},\mathcal{A^*}}) -n \delta  w_{\max}} \label{eqn:qptas_sub_1} \\
& \geq & M_{ {\cal I} }( {\cal A}^* )- n\delta w_{\max} \cdot \sum_{ k \in [K] } \frac{\theta_k }{1+w(C_k^{\mathcal{I}, \mathcal{A}^*})} \nonumber \\
& \geq & M_{ {\cal I} }( {\cal A}^* )- n\delta w_{\max} \nonumber \\
& \geq & \left(1- \frac{2n\delta}{\epsilon}\right) \cdot M_{ {\cal I} }( {\cal A}^* ) \label{eqn:qptas_sub_2} \\
&=& (1-\epsilon)\cdot M_{ {\cal I} }( {\cal A}^* ) \ . \nonumber
\end{eqnarray}
Here, inequality~\eqref{eqn:qptas_sub_1} follows from~\eqref{eqn:tildew_w_cons}, noting that the denominator $1+w(C_k^{\mathcal{I},\mathcal{A^*}}) -n \delta  w_{\max}$ is strictly positive, since $\delta = \frac{\epsilon^2}{2n}$ and $w_{\max} \leq \frac{ 1 }{ \eps }$, implying that $n \delta w_{\max} \leq \frac{\eps}{2} < 1$. To better understand inequality~\eqref{eqn:qptas_sub_2}, note that the optimal market share $M_{ {\cal I} }( {\cal A}^* )$ can be related to $w_{\max}$ by observing that $M_{ {\cal I} }( {\cal A}^* ) \geq \frac{w_{\max}}{1+w_{\max}} \geq \frac{\epsilon}{2} \cdot w_{\max}$, where the first inequality follows from~\eqref{obs_non_triv} and the second inequality holds since $w_{\max} \leq \frac{ 1 }{ \eps }$. 

\subsection{Proof of Lemma~\ref{lem:tildeA_for_I}} 
\label{page:opt_wmax}
\label{app:proof_lem_tildeA_for_I}

Our starting observation is that, for any assignment ${\cal A}$, the stopping point $s^{\mathcal{I}, {\cal A}}_k$ of every customer $k \in [K]$ with respect to ${\cal I}$ cannot occur earlier than her stopping point $s^{\tilde{\mathcal{I}}, {\mathcal{A}}}_k$ with respect to $\tilde{\cal I}$. Indeed, by definition of these points, we have
\begin{eqnarray*}
s^{\mathcal{I}, {\cal A}}_k & = & \min \left\{ p \in [n] :  w( {\cal A}[1,p] ) \geq r_p^k \right\} \\
& \geq & \min \left\{ p \in [n] :  \tilde{w}( {\cal A}[1,p] ) \geq r_p^k \right\} \\
& = & s^{\tilde{\mathcal{I}}, {\mathcal{A}}}_k \ ,
\end{eqnarray*}
where the inequality above holds since $\tilde{w}_i \geq w_i$ for every product $i \in [n]$.

By specializing this observation for the approximate assignment $\tilde{\cal A}$, it follows that $s^{\mathcal{I}, \tilde{\cal A}}_k \geq s^{\tilde{\mathcal{I}}, \tilde{\mathcal{A}}}_k$, implying in turn that the consideration set $C^{\mathcal{I},\tilde{\mathcal{A}}}_k$ of every customer $k \in [K]$ with respect to the original instance ${\cal I}$ contains her consideration set $C^{\tilde{\mathcal{I}},\tilde{\mathcal{A}}}_k$ with respect to the modified instance $\tilde{\cal I}$. Consequently, the expected market share of the assignment $\tilde{\cal A}$ in terms of ${\cal I}$ can be lower-bounded by noticing that
\begin{eqnarray*}
    M_{\mathcal{I}}(\tilde{\mathcal{A}}) & = & \sum_{k \in [K]} \theta_k \cdot \frac{w(C^{\mathcal{I}, \tilde{\mathcal{A}}}_k)}{1 + w(C^{\mathcal{I}, \tilde{\mathcal{A}}}_k)}  \\
    & \geq & \sum_{k \in [K]} \theta_k \cdot \frac{w(C^{\tilde{\mathcal{I}}, \tilde{\mathcal{A}}}_k)}{1 + w(C^{\tilde{\mathcal{I}}, \tilde{\mathcal{A}}}_k)}   \\    
    & \geq & \sum_{k \in [K]} \theta_k \cdot \frac{ \tilde{w} (C^{\tilde{\mathcal{I}}, \tilde{\mathcal{A}}}_k)-n\delta w_{\max}}{1 + \tilde{w} (C^{\tilde{\mathcal{I}}, \tilde{\mathcal{A}}}_k)} \ ,
\end{eqnarray*}
where the last inequality holds since $\tilde{w}_i \leq w_i +\delta w_{\max}$ for every product $i \in [n]$. Further simplifying the latter expression, we have
\begin{eqnarray}
M_{\mathcal{I}}(\tilde{\mathcal{A}}) & \geq & M_{\tilde{\mathcal{I}}}(\tilde{\mathcal{A}}) - n\delta w_{\max} \nonumber\\
    & \geq & (1-\eps)\cdot\opt(\tilde{\mathcal{I}})-n\delta w_{\max} \label{eqn:example_1} \\
    & \geq &
    (1-\epsilon)^2\cdot\opt(\mathcal{I})-\eps\cdot\opt(\mathcal{I}) \label{eqn:example_2} \\
    & \geq & (1-3\epsilon)\cdot \opt(\mathcal{I}) \ . \nonumber
\end{eqnarray}
Here, inequality~\eqref{eqn:example_1} is obtained by recalling that $\tilde{\mathcal{A}}$ is a $(1-\eps)$-approximate assignment for the modified instance $\tilde{\cal I}$. Inequality~\eqref{eqn:example_2} follows from Lemma~\ref{lem:qptas_opt_maintain}, combined with the already-established claim that $n\delta w_{\max}  \leq \eps \cdot \opt (  {\cal I} )$; see last paragraph of Section~\ref{app:proof_qptas_opt_maintain}.

%% file: TEX-PTAS-Overview.tex
\section{Truly Polynomial-Time Approximation Scheme: Technical Overview} \label{sec:overview}

In this section, we provide a high-level overview of our main algorithmic contribution, culminating in a polynomial-time approximation scheme for the market share ranking problem. To clearly mark our objective, we restate the precise performance guarantees of this finding. 

\maintheorem*

\paragraph{Outline.} In Section~\ref{subsec:th_prop}, we describe how to geometrically partition the underlying set of products into logarithmically-many classes according to their preference weights. Based on this partition, we introduce the family of ``sorted-within-class" assignments, where the products in each class are sequentially assigned by weakly-increasing weight order. Interestingly, we prove that at least one such assignment is near-optimal. In Section~\ref{sec_guess_b_w}, we delve deeper into structural properties of the latter assignment, which is further partitioned into blocks of positions, based on their cumulative weight. These blocks will be characterized by a handful of basic statistics, including their number of positions, total weight, and highest-index weight class. In Section~\ref{subsec:b_w_part}, we consider yet another partition of the product set, dubbed as being ``good", where each subset approximates its corresponding block in terms of these statistics. Due to their involved nature, the computational aspects of constructing good partitions will be discussed in Section~\ref{sec:compute_partition}. Sections~\ref{subsec:convert_b_w} and \ref{subsec:ptas_anls} will elaborate on how such a partition can be efficiently converted into a near-optimal position-to-product assignment.

\subsection{Weight classes and sorted-within-class assignments} \label{subsec:th_prop}

As explained in Section~\ref{subsec:qptas}, to devise an approximation scheme for an arbitrarily-structured instance ${\cal I}$ of the market share ranking problem, it suffices to obtain such an approximation for its bounded-ratio counterpart, $\tilde{\mathcal{I}}$, where the extremal weight ratio is polynomial in $n$ and $\frac{1}{\eps}$. We remind the reader that, in this modified instance, the maximal preference weight is $w_{\max} \leq \frac{1}{\eps}$, whereas the minimal such weight is $w_{\min} \geq \frac{\eps^2}{2n} \cdot w_{\max}$. 

\paragraph{Weight classes.} Inspired by the starting point of \citet[Sec.~5.2]{DerakhshanGMM22}, we begin by geometrically partitioning the underlying set of products into a sequence of weight classes ${\cal G}_1, \ldots, {\cal G}_Q$ by powers of $1 + \eps$. Specifically, letting $Q = \lceil\log_{1+\eps}(\frac{2n}{\eps^3}) \rceil  = O( \frac{ 1 }{ \eps } \log \frac{ n }{ \eps })$, each such class is given by
\[ \mathcal{G}_q ~~=~~ \left\{ i \in [n] : w_i \in \left[(1+\eps)^{q-1} \cdot\frac{\eps^2}{2n} \cdot w_{\max}, (1+\eps)^{q}\cdot \frac{\eps^2}{2n} \cdot w_{\max}\right)  \right\} \ .\]
On top of this partition, we define the index set of ``heavy" classes ${\cal Q}_{\myheavy} = \{ q_{\min}, \ldots, Q \}$, where $q_{\min}$ is the minimal index $q$ for which $(1+\eps)^{q-1} \geq 2n$. As such, within these classes, each product has a preference weight of at least $\eps^2 w_{\max}$, and we say that product $i$ is heavy when $i \in \bigcup_{q \in \mathcal{Q}_{\myheavy}}\mathcal{G}_q$. By these definitions, one can easily verify that $|{\cal Q}_{\myheavy}| = O(\frac{1}{\eps} \log \frac{ 1 }{ \eps })$. Any other product will be dubbed as being ``light", with the index set of light weight classes defined as $\mathcal{Q}_{\mylight} =  [Q] \setminus {\cal Q}_{\myheavy}$. 

\paragraph{Sorted-within-class assignments.} Now, let us introduce a special family of assignments, implicitly considered by \cite{DerakhshanGMM22}. To this end, we say that an assignment ${\cal A} : [n] \to [n]$ is sorted-within-class when, for every weight class ${\cal G}_q$, the internal order by which ${\cal A}$ places $\mathcal{G}_q$-products along the sequence of positions is weakly-increasing by weight. In other words, for any pair of products $\{i_1, i_2\} \subseteq {\cal G}_q$ with $w_{i_1} < w_{i_2}$, we have ${\cal A}^{-1}( i_1 ) < {\cal A}^{-1}( i_2 )$. Given this structural restriction, we proceed by considering the following question: How well can we approximate the market share ranking problem by limiting our attention to sorted-within-class assignments? Somewhat hidden within the work of \citeauthor{DerakhshanGMM22}~lies the next claim, showing that at least one such assignment is near-optimal.

\begin{lemma} \label{lemma_a_up}
There exists a sorted-within-class assignment $\mathcal{A}^{\uparrow}$ with an expected market share of  $M( \mathcal{A}^{\uparrow} ) \geq (1 - \eps) \cdot \opt( {\cal I} )$.
\end{lemma}

The reasoning behind this result is that, starting with an optimal assignment ${\cal A}^*$, we can arrive at a sorted-within-class assignment $\mathcal{A}^{\uparrow}$ by keeping the products of each weight class ${\cal G}_q$ within the same set of positions, ${\cal A}^{*-1}( {\cal G}_q )$. However, their internal order will be modified, placing these products in weakly-increasing weight order, thereby creating a sorted-within-class assignment. It is easy to verify that this alteration can only translate the stopping point of any customer $k$ to higher-indexed positions, meaning that $s_k^{ \mathcal{A}^{\uparrow} } \geq s_k^{ \mathcal{A}^* }$, since $w(\mathcal{A}^{\uparrow}[1, s_k^{ \mathcal{A}^*}]) \leq   w(\mathcal{A}^{*}[1, s_k^{ \mathcal{A}^*}])$. In addition, the total weight of her consideration set with respect to $\mathcal{A}^{\uparrow}$ nearly matches its analogous weight with respect to $\mathcal{A}^*$. Indeed,
\begin{eqnarray*}
w(C^{\mathcal{A}^{\uparrow}}_k) & \geq & w(\mathcal{A}^{\uparrow}[1, s^{\mathcal{A}^*}_k]) \\ & \geq & \frac{1}{1+\eps} \cdot w(\mathcal{A}^{*}[1, s^{\mathcal{A}^*}_k])\\  & \geq & 
 (1-\eps) \cdot w(C^{\mathcal{A}^{*}}_k) \ ,
\end{eqnarray*}
where the second inequality holds since by definition, any two products within the same class differ in weight by a factor of at most $1+\eps$. As a result,
\[M(\mathcal{A}^{\uparrow}) ~~=~~ \sum_{k \in [K]}\theta_k \cdot \frac{w(C^{\mathcal{A}^{\uparrow}}_k)}{1 + w(C^{\mathcal{A}^{\uparrow}}_k)}  ~~\geq~~ (1-\eps)\cdot\sum_{k \in [K]} \theta_k \cdot \frac{ w(C^{\mathcal{A}^{*}}_k)}{1 + w(C^{\mathcal{A}^{*}}_k)} ~~=~~ (1-\eps)\cdot M(\mathcal{A}^*) \ . \]

\subsection{The block structure of \bstitle{\mathcal{A}^\uparrow}}\label{subsec:block_bw}

In what follows, we flesh out important structural properties of the sorted-within-class assignment $\mathcal{A}^{\uparrow}$, whose existence has been established in Lemma~\ref{lemma_a_up}. In a nutshell, we will explain how to partition the sequence of positions $1, \ldots, n$ into contiguous blocks, based on their cumulative weight. These blocks will be associated with several summary statistics, including their number of positions, total weight, and highest-index weight class. We mention in passing that, from an algorithmic standpoint, the assignment $\mathcal{A}^{\uparrow}$ is obviously unknown, meaning that the upcoming discussion is still analytical in nature.

\paragraph{Partition into blocks.} With respect to the assignment $\mathcal{A}^{\uparrow}$, we move on to introduce its corresponding sequence of  blocks $\mathcal{B}^{\uparrow}_0, \ldots, \mathcal{B}^{\uparrow}_L, \mathcal{B}^{\uparrow}_\infty$, noting that each such block is simply a set of successive positions. These blocks are sequentially defined as follows:
\begin{itemize}
    \item Our first block $\mathcal{B}^{\uparrow}_0$ is given by the prefix of positions in which the cumulative total weight remains strictly under $\eps^3 w_{\max}$. In other words, $\mathcal{B}^{\uparrow}_0 = [1,p_0]$, where $p_0$ is the maximal position $p \in [n]$ for which $w(\mathcal{A}^{\uparrow}[1,p]) < \eps^3 w_{\max}$.  It is worth mentioning that when $w(\mathcal{A}^{\uparrow}(1)) \geq \eps^3 w_{\max}$, such a position does not exist, in which case we end up with an empty block. In addition, this definition implies that $\mathcal{B}^{\uparrow}_0$ does not contain any heavy product, since each such product by itself has a preference weight of at least $\eps^2 w_{\max}$.

    \item The next block $\mathcal{B}^{\uparrow}_1$ begins at position $|\mathcal{B}^{\uparrow}_0|+1$, stretching as long as the cumulative total weight (including $\mathcal{B}^{\uparrow}_0$) remains strictly under $(1+\eps)\cdot \eps^3 w_{\max}$. That is, $\mathcal{B}^{\uparrow}_1 = [| \mathcal{B}^{\uparrow}_0 |+1,p_1]$, where $p_1$ is the maximal position $p \in [n]$ for which $w(\mathcal{A}^{\uparrow}[1,p]) < (1+\eps)\cdot \eps^3 w_{\max}$. As before, this block could be empty.
    
    \item In general, given that we have already defined the sequence $\mathcal{B}^{\uparrow}_0, \ldots, \mathcal{B}^{\uparrow}_{\ell - 1}$, the next block $\mathcal{B}^{\uparrow}_\ell$ will begin at position $| \mathcal{B}^{\uparrow}_0| + \cdots + | \mathcal{B}^{\uparrow}_{\ell - 1} | + 1$, stretching up to and including position $p_\ell$, which is the maximal position $p \in [n]$ for which $w(\mathcal{A}^{\uparrow}[1,p]) < (1+\eps)^{\ell} \cdot \eps^3 w_{\max}$. This procedure continues up until arriving at block $\mathcal{B}^{\uparrow}_L$, where $L$ is the smallest index $\ell$ for which $(1+\eps)^\ell \cdot \eps^3 > \frac{1}{\eps}$, meaning that $L = \lceil \log_{1+\eps} \left(\frac{1}{\eps^{4}}\right) \rceil = O(\frac{1}{\eps} \log \frac{1}{\eps} )$. 

    \item In light of these definitions, for any position $p$ beyond block $\mathcal{B}^{\uparrow}_L$, we hit a cumulative total weight of $w(\mathcal{A}^{\uparrow}[1,p]) \geq (1+\eps)^L \cdot \eps^3 w_{\max} > \frac{ w_{\max} }{ \eps }$. These positions will be packed into our final block, denoted by $\mathcal{B}^{\uparrow}_\infty = [ \sum_{\ell \in [L]} | \mathcal{B}^{\uparrow}_{\ell}| + 1, n]$.
\end{itemize}

\paragraph{Block statistics.}\label{sec_guess_b_w} To avoid cumbersome notation, for every $\ell \in [L]$, we make use of $\mathcal{W}^{\uparrow}_\ell = w( \mathcal{A}^{\uparrow}( \mathcal{B}^{\uparrow}_\ell ) )$ to designate the combined weight of the products placed by the assignment $\mathcal{A}^{\uparrow}$ within block $\mathcal{B}^{\uparrow}_\ell$. In addition, $\beta^{\uparrow}_\ell =  | \mathcal{B}^{\uparrow}_\ell |$ stands for the size of this block, i.e., its number of positions. Finally, we use $q^{\uparrow}_{\ell}$ to denote the highest-index weight class from which at least one product appears in $\mathcal{B}^{\uparrow}_\ell$, namely, $q^{\uparrow}_{\ell} = \max \{ q \in [Q] : \mathcal{A}^{\uparrow}({\mathcal{B}^{\uparrow}_{\ell}) \cap \mathcal{G}_q \neq \emptyset}\}$. Once again, it is important to emphasize that, from an algorithmic perspective, these quantities are obviously unknown.

\subsection{Computing good partitions}
\label{subsec:b_w_part}

\paragraph{Good partitions.}\label{sec_good_part} We proceed by defining the notion of a ``good partition", which is a partition $\mathcal{S} = ({\cal S}_1, \ldots {\cal S}_L, {\cal S}_{\infty})$ of the product set that satisfies the next four properties:
\begin{enumerate}
    \item \label{good_assign_1}{\em Bounded size}: Each set $\mathcal{S}_\ell$ contains at most $\beta^{\uparrow}_\ell$ products, i.e., $|\mathcal{S}_\ell| \leq \beta^{\uparrow}_\ell$ for every $\ell \in [L]$. 
    
    \item \label{good_assign_2}{\em Bounded weight}: The total weight of the products within each set $\mathcal{S}_\ell$ nearly matches ${\mathcal{W}}^{\uparrow}_\ell$, specifically meaning that $(1-\eps)\cdot{\mathcal{W}}^{\uparrow}_\ell - \eps^4 w_{\max} \leq w(\mathcal{S}_\ell) \leq {\mathcal{W}}^{\uparrow}_\ell$ for every $\ell \in [L]$.
    
    \item \label{good_assign_3}{\em Highest-index weight class}: For every $\ell \in [L]$ with  $q^{\uparrow}_\ell \in {\cal Q}_{\myheavy}$, the set $\mathcal{S}_{\ell}$ contains at least one product from $\mathcal{G}_{q^{\uparrow}_\ell}$, that is, $\mathcal{S}_\ell \cap \mathcal{G}_{q^{\uparrow}_\ell} \neq \emptyset$.

    \item \label{good_assign_4}{\em Prefix subsets}: The collection of products in $\mathcal{S}_{1}, \ldots, \mathcal{S}_L $ is a subset of those assigned by $\mathcal{A}^{\uparrow}$ to the blocks  $\mathcal{B}^{\uparrow}_0, \ldots, \mathcal{B}^{\uparrow}_L$, meaning that $\bigcup_{\ell \in [L]} \mathcal{S}_{\ell} \subseteq  \bigcup_{\ell \in [L]_0} \mathcal{A}^{\uparrow}(\mathcal{B}^{\uparrow}_{\ell}) $.
\end{enumerate}
In summary, these properties make use of the basic statistics characterizing $\mathcal{B}^{\uparrow}_{0}, \ldots, \mathcal{B}^{\uparrow}_{L}, \mathcal{B}^{\uparrow}_{\infty}$, as defined in Section~\ref{sec_guess_b_w}, thereby relating each subset $\mathcal{S}_{\ell}$ to its corresponding block $\mathcal{B}^{\uparrow}_{\ell}$ in terms of size, total weight, and highest-index weight class. We mention in passing that, even though our sequence of blocks is $\mathcal{B}^{\uparrow}_0, \ldots, \mathcal{B}^{\uparrow}_L, \mathcal{B}^{\uparrow}_\infty$, good partitions do not include a set $\mathcal{S}_0$ corresponding to  $\mathcal{B}^{\uparrow}_0$.

 \paragraph{Motivating good partitions.} At this point in time, it might be unclear why such partitions are useful. To intuitively address this question, we remind the reader that the original solution space of market share ranking consists of position-to-product assignments, i.e., linear orderings of the product set; in contrast, partitions are collections of pairwise-disjoint sets. Thinking in algorithmic terms, good partitions appear to be easier to deal with in comparison to directly interacting with near-optimal assignments, since the former objects are represented as a collection of unordered sets, as opposed to the very rigid structure of an assignment. On the other hand, we insist on attaining properties \ref{good_assign_1}-\ref{good_assign_4}, which seem to involve an additional layer of complexity. As shown in the sequel, the specific structural properties of good partitions allow us to efficiently create assignments that are very much mimicking the unknown near-optimal assignment $\mathcal{A}^{\uparrow}$, and are therefore nearly-optimal themselves.

\paragraph{Constructing good partitions.} Rather than identifying a single good partition, it will be convenient to construct a family of candidate partitions, $\mathcal{F}$, arguing that it contains at least one good partition. A family $\mathcal{F}$ meeting this criterion will be referred to as a ``good family''. In Section~\ref{sec:compute_partition}, we devise efficient enumeration ideas, employed in order to jointly  guess selected block statistics. Based on these guesses, we will describe our algorithmic approach for assigning products to each of the subsets $\mathcal{S}_1, \ldots, \mathcal{S}_L, \mathcal{S}_\infty$, thus creating a candidate partition for each guess. Eventually, these ideas will be shown to produce a good family, arguing that for any fixed $\eps >0$, the running time of this approach is polynomial in $n$, as formally stated below.

\begin{theorem}\label{thm:bw_part}
We can construct a good family ${\cal F}$ consisting of  $O(n^{O(\frac{1}{\eps^{6}}\log\frac{1}{\eps})})$ partitions. Our construction can be implemented in $O(n^{O(\frac{1}{\eps^{6}}\log\frac{1}{\eps})})$ time.
\end{theorem} 
 
\subsection{Translating  good partitions to approximate assignments}
\label{subsec:convert_b_w}

Assuming to have obtained a good family $\mathcal{F}$ via Theorem~\ref{thm:bw_part}, let us focus on one good partition $\mathcal{S} = (\mathcal{S}_1, \ldots, \mathcal{S}_L, \mathcal{S}_{\infty}) \in \mathcal{F}$. In what follows, we explain how the latter will be converted to an approximate assignment $\tilde{\mathcal{A}}$, noting that this procedure is repeated for all partitions in $\mathcal{F}$, eventually returning the most profitable one. At a high level, $\tilde{\cal A}$ is constructed by judiciously placing the products within $\mathcal{S}_{1}, \ldots, \mathcal{S}_{L}, \mathcal{S}_{\infty}$ one after the other into the positions $1, \ldots, n$. The internal order for each subset is rather arbitrary, with two important exceptions that will be described below. Technically speaking, the assignment $\tilde{\mathcal{A}}$ is constructed as follows:
\begin{itemize}
    \item First, the products in $\mathcal{S}_1$ will be placed in positions $1, \ldots, |\mathcal{S}_1|$, jointly forming the block $\tilde{\cal B}_1$. The internal order between these products will be arbitrary, except for one special case. As mentioned is Section~\ref{sec_guess_b_w},  $q^{\uparrow}_1$ is the largest index $q$ for which $\mathcal{A}^{\uparrow}(\mathcal{B}^{\uparrow}_1) \cap \mathcal{G}_q \neq \emptyset$. Then, when $q^{\uparrow}_1 \in {\cal Q}_{\myheavy}$, the assignment $\tilde{\cal A}$ intentionally places a product from $\mathcal{S}_1 \cap \mathcal{G}_{q^{\uparrow}_1}$ at position $1$, noting that $\mathcal{S}_1 \cap \mathcal{G}_{q^{\uparrow}_1} \neq \emptyset$ by property~\ref{good_assign_3} of good partitions. 

    \item Next, the products in $\mathcal{S}_2$ will be placed in positions $|\mathcal{S}_1|+1, \ldots, |\mathcal{S}_1|+|\mathcal{S}_2|$, dubbing this sequence of positions as block $\tilde{\cal B}_2$. Once again, the internal order between these products is arbitrary, except for the case where $q^{\uparrow}_2 \in {\cal Q}_{\myheavy}$. As before, the assignment $\tilde{\mathcal{A}}$ takes one of the products in $\mathcal{S}_2 \cap {\cal G}_{ q^{\uparrow}_2 }$ and places it at position $| {\cal S}_1 | + 1$. Again, by property~\ref{good_assign_3}, we know that $\mathcal{S}_2 \cap \mathcal{G}_{q^{\uparrow}_2} \neq \emptyset$. 
    
    \item So on and so forth, where in general, for every $\ell \in [L]$, the products in $\mathcal{S}_{\ell}$ will be similarly placed in positions $|\mathcal{S}_1| + \cdots + |\mathcal{S}_{\ell-1}| + 1, \ldots, |\mathcal{S}_1| + \cdots + |\mathcal{S}_{\ell}|$, forming the block $\tilde{\cal B}_{\ell}$. 
    
    \item Finally, the products in $\mathcal{S}_{\infty}$ will be placed in positions $\sum_{ \ell \in [L]}|\mathcal{S}_{\ell}|+1, \ldots, n$, which will be comprising our last block, $\tilde{\cal B}_{\infty}$. It is important to point out that, unlike earlier blocks, the internal order within $\tilde{\cal B}_{\infty}$ will be by weakly-decreasing weights.
\end{itemize}

\subsection{Analysis}\label{subsec:ptas_anls}

In what follows, we conclude the proof of Theorem~\ref{thm:main_result}, arguing that the market share ranking problem admits an $O(n^{O(\frac{1} {\eps^{6}}\log\frac{1}{\eps})})$-time approximation scheme. To this end, we claim that the assignment $\tilde{\cal A}$, as constructed in Section~\ref{subsec:convert_b_w}, guarantees a near-optimal expected market share. Specifically, in the remainder of this section, we establish the next result. 

\begin{lemma} \label{lem:ptas_opt_maintain}
    $M(\tilde{\mathcal{A}}) \geq (1-13\eps)\cdot\opt(\mathcal{I})$.
\end{lemma}

\paragraph{Customer types by stopping points.} To derive this bound, we begin by classifying our collection of customers into three types --- early, midway, and late --- based on their stopping points $\{s^{\mathcal{A}^{\uparrow}}_k\}_{k \in [K]}$ with respect to the sorted-within-class assignment $\mathcal{A}^{\uparrow}$. These types are characterized by the specific block where $s^{\mathcal{A}^{\uparrow}}_k$ is located, along the following trichotomy: 
\begin{itemize}
    \item {\em Early stoppers}: Our first type $\Tearly$ consists of customers $k \in [K]$ whose stopping point $s^{\mathcal{A}^{\uparrow}}_k$ resides within block $\mathcal{B}^{\uparrow}_0$. 
    
    \item {\em Midway stoppers}: Next, the type $\Tmid$ consists of customers $k \in [K]$ whose stopping point resides within one of the blocks $\mathcal{B}^{\uparrow}_1, \ldots, \mathcal{B}^{\uparrow}_L$. In other words, $s^{\mathcal{A}^{\uparrow}}_k \in \mathcal{B}^{\uparrow}_{[1, L]}$, where by convention, $\mathcal{B}^{\uparrow}_{[1, L]} = \bigcup_{\ell \in [L]} \mathcal{B}^{\uparrow}_{\ell}$. 
    
    \item {\em Late stoppers}: Finally, $\Tlate$ is comprised of customers $k \in [K]$ whose stopping point $s^{\mathcal{A}^{\uparrow}}_k$ resides within block $\mathcal{B}^{\uparrow}_{\infty}$.
\end{itemize}

\paragraph{Type-dependent market share bounds.} Depending on the above-mentioned types, we proceed by relating between the market share $M_k( \tilde{\cal A} )$ of each customer $k$ with respect to the approximate assignment $\tilde{\cal A}$ and her analogous market share $M_k( \mathcal{A}^{\uparrow} )$ with respect to the assignment $\mathcal{A}^{\uparrow}$: 

\begin{itemize}
    \item {\em Early stoppers.} Starting with early stoppers, as mentioned in Section~\ref{subsec:b_w_part}, good partitions do not include a set corresponding to $\mathcal{B}^{\uparrow}_0$, with the customers stopping within this block forming $\Tearly$. However, in the next claim, whose proof appears in Appendix~\ref{app:proof_tearly}, we show that the contribution of any such customer to the overall market share is insignificant.

    \begin{lemma} \label{lem:market_share_early}
    $M_k ( {\cal A}^{ \uparrow }  ) \leq \eps^3 w_{\max}$, for every $k \in \Tearly$.  
    \end{lemma}
    
    \item {\em Midway stoppers.} In this case, we take advantage of how good partitions are defined to show that, for every customer $k \in \Tmid$, her expected market share $M_k(\tilde{\cal A})$ closely approximates $M_k({\cal A}^{\uparrow})$. The proof of this result is provided in Appendix~\ref{app:proof_tmid}.

    \begin{lemma} \label{lem:market_share_mid}
    $M_k ( \tilde{\cal A} ) \geq (1-2\eps) \cdot M_k({\cal A}^{\uparrow}) -4 \eps^2  w_{\max}$, for every $k \in \Tmid$.  
    \end{lemma}
    
    \item {\em Late  stoppers.} Finally, we exploit the unique structure of good partitions to argue that, for every customer $k \in \Tlate$, her expected market share $M_k(\tilde{\cal A})$ nearly matches $M_k({\cal A}^{\uparrow})$, noting that the arguments involved will be substantially different from those employed for midway stoppers. This result is established in Appendix~\ref{app:proof_tlate}.

    \begin{lemma} \label{lem:market_share_late}
    $M_k ( \tilde{\cal A} ) \geq (1-2\eps) \cdot M_k({\cal A}^{\uparrow}) -4 \eps^2  w_{\max}$, for every $k \in \Tlate$.  
    \end{lemma}
\end{itemize}

\paragraph{Putting everything together.} We are now ready to conclude the proof of Lemma~\ref{lem:ptas_opt_maintain}, claiming that $\tilde{\cal A}$ is near-optimal. Given the above-mentioned performance guarantees for each customer type, we proceed to compare the expected market share of our candidate assignment $\tilde{\mathcal{A}}$ to that of the optimal assignment $\mathcal{A}^*$, by observing that
\begin{eqnarray}
M(\tilde{\mathcal{A}}) & = & \nonumber \sum_{k \in [K]} \theta_k \cdot M_k( \tilde{\mathcal{A}} ) \\
& \geq & \label{ptas_opt_1} \sum_{k \in \Tmid \cup \Tlate} \theta_k \cdot  \left( (1-2\eps) \cdot M_k({\cal A}^{\uparrow}) -4\eps^2  w_{\max}\right) \\ 
& \geq &  (1-2\eps) \cdot M(\mathcal{A}^{\uparrow}) - \sum_{k \in \Tearly} \theta_k \cdot M_k( {\cal A}^{\uparrow} ) - 4\eps^2   w_{\max} \nonumber\\
& \geq & \label{ptas_opt_2} (1-2\eps) \cdot M(\mathcal{A}^{\uparrow}) - 5\eps^2   w_{\max} \\
& \geq & \label{ptas_opt_3} (1-2\eps)(1-\eps) \cdot \opt(\mathcal{I}) - 5\eps^2   w_{\max} \\
& \geq & \label{ptas_opt_4} (1-13\eps) \cdot \opt(\mathcal{I}) \ .
\end{eqnarray}
Here, inequality~\eqref{ptas_opt_1} follows from Lemmas \ref{lem:market_share_mid} and \ref{lem:market_share_late}, which lower-bound the  expected market share of midway and late stoppers. Inequality~\eqref{ptas_opt_2} is obtained by plugging in Lemma~\ref{lem:market_share_early}, which upper-bounds the expected market share of early stoppers. Inequality~\eqref{ptas_opt_3} follows from Lemma~\ref{lemma_a_up}, stating that $M(\mathcal{A}^{\uparrow}) \geq (1-\eps) \cdot \opt(\mathcal{I})$. Finally, to arrive at inequality~\eqref{ptas_opt_4}, we note that $\opt(\mathcal{I}) \geq \frac{\eps}{2} \cdot w_{\max}$, as observed while proving Lemma~\ref{lem:qptas_opt_maintain} (see page~\pageref{page:opt_wmax}).

%% file: TEX-Partition.tex
\section{Computing  Good Partitions} \label{sec:compute_partition}

In this section, we present an enumeration-based procedure for constructing a good family of $O(n^{O(\frac{1}{\eps^{6}}\log\frac{1}{\eps})})$ partitions, as formally stated in Theorem~\ref{thm:bw_part}. At a high level, our approach proceeds in two steps, initially assigning heavy products to the sets $\mathcal{S}_1, \ldots, \mathcal{S}_L, \mathcal{S}_\infty$, and subsequently deciding on the assignment of light products, with distinct allocation rules for each case. Toward this objective, in Section~\ref{sec_block_guess}, we develop an efficient guessing procedure for the number of products appearing in each block out of each heavy weight class, as well as for the cumulative weight of light products assigned to each such block. Section~\ref{sec_asgn_heavy} explains how to utilize these guesses for assigning heavy products to the sets $\mathcal{S}_{1}, \ldots, \mathcal{S}_{L}, \mathcal{S}_\infty$. In Section~\ref{sec_asgn_light}, we describe how light products are assigned, ensuring that their cumulative weights are aligned with our previously-obtained guesses. Finally, Sections~\ref{sec_ptas_analysis} and~\ref{sec_ptas_analysis_cont} analyze this construction, proving that it indeed yields a good family of $O(n^{O(\frac{1}{\eps^{6}}\log\frac{1}{\eps})})$ partitions.

\subsection{Guessing block properties}\label{sec_block_guess}
We begin by describing an efficient guessing procedure for two important quantities, related to the block statistics of the assignment $\mathcal{A}^{\uparrow}$, whose finer details were discussed in Section~\ref{sec_guess_b_w}. Specifically, we first guess the number of products assigned to each block $\mathcal{B}^{\uparrow}_{\ell}$ from each of the weight classes $\{\mathcal{G}_q\}_{q \in \mathcal{Q}_{\myheavy}}$. Subsequently, we proceed by obtaining an under-estimate for the combined weight of the products across all weight classes $\{\mathcal{G}_q\}_{q \in \mathcal{Q}_{\mylight}}$ assigned to each block. 

\paragraph{Step 1: Guessing heavy-class assignment quantities.} For every $\ell \in [L]$ and $q \in \mathcal{Q}_{\myheavy}$, let $\beta^{\uparrow}_{\ell, q}$ be the number of products assigned by $\mathcal{A}^{\uparrow}$ from weight class $\mathcal{G}_q$ to block $\mathcal{B}^{\uparrow}_{\ell}$, i.e., $\beta^{\uparrow}_{\ell, q} =|\mathcal{A}^{\uparrow}(\mathcal{B}^{\uparrow}_{\ell}) \cap \mathcal{G}_q|$. In addition, let  $\beta^{\uparrow}_{\ell, \mylight} = |\mathcal{A}^{\uparrow}(\mathcal{B}^{\uparrow}_{\ell}) \cap (\bigcup_{q \in \mathcal{Q}_{\mylight}} \mathcal{G}_q)|$ be the number of light products assigned to $\mathcal{B}^{\uparrow}_{\ell}$. Altogether, these quantities provide us with the exact size $\beta^{\uparrow}_{\ell}$ of block $\mathcal{B}^{\uparrow}_{\ell}$, since $\beta^{\uparrow}_{\ell} = \sum_{q \in \mathcal{Q}_{\myheavy}} \beta^{\uparrow}_{\ell, q} + \beta^{\uparrow}_{\ell, \mylight}$.
Moreover, recalling that ${q}^{\uparrow}_{\ell} = \max\{q \in [Q] : \beta^{\uparrow}_{\ell, q} \geq 1\}$ is the highest-index weight class from which at least one product appears in $\mathcal{B}^{\uparrow}_\ell$, this parameter can be inferred whenever it corresponds to a heavy class.

Our objective is to construct a family of guesses, $\mathcal{F}_{\beta}$, each taking the form $\{\hat{\beta}_{\ell, q}\}_{\ell \in [L], q \in \mathcal{Q}_{\myheavy}\cup \{\mylight\}}$, such that $\{{\beta}^{\uparrow}_{\ell, q}\}_{\ell \in [L], q \in \mathcal{Q}_{\myheavy}\cup \{\mylight\}}$ is necessarily a member of $\mathcal{F}_{\beta}$. To this end, there are only $O(n)$ guesses to consider for each $\beta^{\uparrow}_{\ell, q}$. Noting that $L \cdot (|\mathcal{Q}_{\myheavy}| + 1)$ such guesses are required, the total number of joint values that could be created is $O(n^{O(|\mathcal{Q}_{\myheavy}|\cdot L)}) =  O({n}^{O(\frac{1}{\eps^2}\log^2(\frac{1}{\eps}))}) $, since $|\mathcal{Q}_{\myheavy}| = O(\frac{1}{\eps} \log \frac{1}{\eps} )$ and $L = O(\frac{1}{\eps} \log \frac{1}{\eps} )$, as explained in Sections~\ref{subsec:th_prop} and \ref{subsec:block_bw}, respectively.

\paragraph{Step 2: Estimating light-class assignment weights.} Next, for every $\ell \in [L]$ and $q \in \mathcal{Q}_{\mylight}$, let ${\cal W}^{\uparrow}_{\ell, q} = w( \mathcal{A}^{\uparrow}( \mathcal{B}^{\uparrow}_\ell ) \cap \mathcal{G}_q)$ represent the combined weight of the products from class $\mathcal{G}_q$ assigned by $\mathcal{A}^{\uparrow}$ to block $\mathcal{B}^{\uparrow}_{\ell}$. In Appendix~\ref{app_clm_guess_w}, we devise a guessing procedure for creating a family $\mathcal{F}_W$ of $O(n^{O({\frac{1}{\eps^{6}}\log\frac{1}{\eps})}})$ estimates, each being of the form $\{\hat{\cal W}_{\ell, q}\}_{\ell \in [L], q \in \mathcal{Q}_{\mylight}}$. Our construction guarantees that there exists at least one estimate in $\mathcal{F}_W$ such that, for every $\ell \in [L]$ and $q \in \mathcal{Q}_{\mylight}$, we have
\begin{eqnarray}\label{eqn_w_property}
    \mathcal{W}^{\uparrow}_{\ell, q} -\frac{\eps^4 w_{\max}}{|\mathcal{Q}|} ~~\leq~~ \hat{\cal W}_{\ell, q} ~~\leq~~ \mathcal{W}^{\uparrow}_{\ell, q} \ . 
\end{eqnarray}

\subsection{Assigning heavy products} \label{sec_asgn_heavy}
As our next step in constructing a good partition $\mathcal{S} = (\mathcal{S}_1, \ldots, \mathcal{S}_L, \mathcal{S}_\infty)$, we begin by describing how heavy products will be assigned to each of these sets. Given our guesses for $\{{\beta}^{\uparrow}_{\ell, q}\}_{\ell \in [L], q \in \mathcal{Q}_{\myheavy}}$, attained in step 1, we know exactly how many products were assigned by $\mathcal{A}^{\uparrow}$ from each heavy weight class to each of the blocks $\mathcal{B}^{\uparrow}_{1}, \ldots, \mathcal{B}^{\uparrow}_{L}$; however, we do not know the exact identities of these products. Consequently, our procedure for assigning heavy products to $\mathcal{S}_{1}, \ldots, \mathcal{S}_{L}, \mathcal{S}_{\infty}$ proceeds as follows:

\begin{itemize}
    \item For every $q \in \mathcal{Q}_{\myheavy}$, let us index the products in weight class $\mathcal{G}_q$ as $\{ \langle u, q \rangle \}_{u \in [|\mathcal{G}_q|]}$, by order of weakly-increasing weight, i.e., $w_{\langle 1, q \rangle} \leq \cdots \leq w_{\langle |\mathcal{G}_q|, q \rangle}$. 
    In addition, for every $u^- \leq u^+$, let ${\cal G}_q[u^{-}, u^{+}] = \{\langle u^{-}, q \rangle, \ldots, \langle u^{+}, q \rangle\}$ be the set of $u^-$-th to $u^+$-th products in this sequence.
    \item Now, for every $q \in \mathcal{Q}_{\myheavy}$, products from weight class $\mathcal{G}_q$ will be partitioned between $\mathcal{S}_1, \ldots, \mathcal{S}_L, \mathcal{S}_\infty$ as follows:
        \begin{itemize}
            \item {\em Assignment to $\mathcal{S}_1$}: The lowest $\beta^{\uparrow}_{1,q}$-ranked products in $\mathcal{G}_q$, corresponding to ${\cal G}_{q}[1, \beta^{\uparrow}_{1, q}]$, are assigned to $\mathcal{S}_1$.
            \item {\em Assignment to $\mathcal{S}_2$}: Then, $\mathcal{S}_2$ will be assigned with the next $\beta^{\uparrow}_{2, q}$-ranked products, corresponding to ${\cal G}_q[\beta^{\uparrow}_{1, q}+1, \beta^{\uparrow}_{1, q} + \beta^{\uparrow}_{2, q}]$.
            \item {\em Assignment to $\mathcal{S}_3, \ldots, \mathcal{S}_L$}: We proceed along these lines for each $3 \leq \ell\leq L$ by assigning $\mathcal{G}_q[\sum_{\hat{\ell} \leq \ell-1}\beta^{\uparrow}_{\hat{\ell},q}+1, \sum_{\hat{\ell}\leq \ell}\beta^{\uparrow}_{\hat{\ell},q}]$ to $\mathcal{S}_{\ell}$. This procedure is clearly well-defined, since $\sum_{\ell \in [L]}\beta^{\uparrow}_{\ell, q} = |\mathcal{G}_q \cap \mathcal{A}^{\uparrow}(\mathcal{B}^{\uparrow}_{[1, L]})| \leq |\mathcal{G}_q|$.
            \item {\em Assignment to $\mathcal{S}_\infty$}: Finally, we assign the remaining products $\mathcal{G}_q[\sum_{{\ell} \leq L}\beta^{\uparrow}_{{\ell},q}+1, |\mathcal{G}_q|]$ to $\mathcal{S}_{\infty}$.
        \end{itemize}
\end{itemize}

\subsection{Assigning light products}\label{sec_asgn_light}
As our last step, we explain how the collection of light products, $\bigcup_{q \in \mathcal{Q}_{\mylight}}\mathcal{G}_{q}$, will be assigned to $\mathcal{S}_1, \ldots, \mathcal{S}_L, \mathcal{S}_\infty$, relying on the estimates $\{\hat{\cal W}_{\ell, q}\}_{\ell \in [L], q \in \mathcal{Q}_{\mylight}}$ for $\{{\cal W}^{\uparrow}_{\ell, q}\}_{\ell \in [L], q \in \mathcal{Q}_{\mylight}}$, obtained in step 2. Similarly to our notation for heavy products, those in each light weight class $\mathcal{G}_q$ will be indexed as $\{\langle u, q \rangle\}_{u \in [|\mathcal{G}_q|]}$ by order of weakly-increasing weight. With this indexing scheme, for every $q \in \mathcal{Q}_{\mylight}$, products from weight class $\mathcal{G}_q$ will be partitioned between $\mathcal{S}_1, \ldots, \mathcal{S}_L, \mathcal{S}_\infty$ as follows:
\begin{itemize}
    \item {\em Assignment to $\mathcal{S}_1$}: First, let $\hat{n}_{1,q}$ be the minimal integer satisfying $w(\mathcal{G}_q[1, \hat{n}_{1,q}]) \geq (1-\eps) \cdot \hat{\cal W}_{1, q}$. To verify that $\hat{n}_{1,q}$ is well-defined, note that we must have $\hat{n}_{1,q} \leq |\mathcal{A}^{\uparrow}(\mathcal{B}^{\uparrow}_1) \cap \mathcal{G}_q|$, since
    \begin{eqnarray}
        w(\mathcal{G}_q[1,|\mathcal{A}^{\uparrow}(\mathcal{B}^{\uparrow}_1) \cap \mathcal{G}_q|]) &\geq&
        |\mathcal{A}^{\uparrow}(\mathcal{B}^{\uparrow}_1) \cap \mathcal{G}_q| \cdot (1+\eps)^{q-1}\cdot \frac{\eps^2}{2n}\cdot w_{\max} \label{light_assign_1} \\
        &\geq &\frac{1}{1+\eps} \cdot w(\mathcal{A}^{\uparrow}(\mathcal{B}^{\uparrow}_1) \cap \mathcal{G}_q)  \label{light_assign_2}\\
        & = & \frac{1}{1+\eps} \cdot \mathcal{W}^{\uparrow}_{1,q} \nonumber \\ 
        &\geq & (1-\eps)\cdot \hat{\cal W}_{1,q} \label{light_assign_3} \ . 
    \end{eqnarray}
    Here, inequalities~\eqref{light_assign_1} and \eqref{light_assign_2} hold since $w_i \in [(1+\eps)^{q-1} \cdot\frac{\eps^2}{2n} \cdot w_{\max}, (1+\eps)^{q}\cdot \frac{\eps^2}{2n} \cdot w_{\max})$ for every product $i \in \mathcal{G}_q$. Inequality~\eqref{light_assign_3} utilizes the right inequality in \eqref{eqn_w_property}, stating that $\hat{\cal W}_{\ell,q} \leq {\cal W}^{\uparrow}_{\ell,q}$ for every $\ell \in [L]$ and $q \in \mathcal{Q}_{\mylight}$. With this observation, the lowest $\hat{n}_{1,q}$-ranked products in $\mathcal{G}_q$, corresponding to $\mathcal{G}_q[1, \hat{n}_{1,q}]$, are assigned to $\mathcal{S}_1$. When $\hat{\cal W}_{1,q} = 0$, no products from $\mathcal{G}_q$ will be assigned to $\mathcal{S}_1$.
    \item {\em Assignment to $\mathcal{S}_2$}: Next, let $\hat{n}_{2,q}$ be the minimal integer satisfying $w(\mathcal{G}_q[\hat{n}_{1,q} + 1, \hat{n}_{1,q} + \hat{n}_{2,q}])  \geq  (1-\eps) \cdot \hat{\cal W}_{2, q}$. Again, to verify that $\hat{n}_{2,q}$ is well-defined, by following arguments similar to those of the previous item, and recalling that $\hat{n}_{1,q} \leq |\mathcal{A}^{\uparrow}(\mathcal{B}^{\uparrow}_1) \cap \mathcal{G}_q|$,
    \begin{eqnarray*}
        w(\mathcal{G}_q[\hat{n}_{1,q} + 1, \hat{n}_{1,q} + |\mathcal{A}^{\uparrow}(\mathcal{B}^{\uparrow}_{2}) \cap \mathcal{G}_q|]) &\geq&
         |\mathcal{A}^{\uparrow}(\mathcal{B}^{\uparrow}_2) \cap \mathcal{G}_q| \cdot (1+\eps)^{q-1}\cdot \frac{\eps^2}{2n}\cdot w_{\max}  \\
        &\geq &\frac{1}{1+\eps} \cdot w(\mathcal{A}^{\uparrow}(\mathcal{B}^{\uparrow}_2) \cap \mathcal{G}_q)  \\
        & = & \frac{1}{1+\eps} \cdot \mathcal{W}^{\uparrow}_{2,q}  \\ 
        &\geq & (1-\eps)\cdot \hat{\cal W}_{2,q} \ , 
    \end{eqnarray*}
    which implies that $\hat{n}_{2,q} \leq |\mathcal{A}^{\uparrow}(\mathcal{B}^{\uparrow}_{2}) \cap \mathcal{G}_q|$. Then, the products in $\mathcal{G}_q[\hat{n}_{1,q}+1, \hat{n}_{1,q} + \hat{n}_{2,q}]$ are assigned to $\mathcal{S}_{2}$.
    \item {\em Assignment to $\mathcal{S}_3, \ldots, \mathcal{S}_L$}: Generally speaking, having already handled $\mathcal{S}_1, \ldots, \mathcal{S}_{\ell-1}$, we define $\hat{n}_{\ell,q}$ as the minimal integer satisfying $w(\mathcal{G}_q[\sum_{\hat{\ell} \leq \ell-1}\hat{n}_{\hat{\ell},q}+1, \sum_{\hat{\ell}\leq \ell }\hat{n}_{\hat{\ell},q}]) \geq  (1-\eps) \cdot \hat{\cal W}_{\ell, q}$. Once again, $\hat{n}_{{\ell},q}$ is well-defined, based on arguments similar to those provided when dealing with $\mathcal{S}_1$ and $\mathcal{S}_2$. Then, we assign ${\cal G}_q[\sum_{\hat{\ell}\leq \ell-1}\hat{n}_{\hat{\ell},q}+1, \sum_{\hat{\ell} \leq \ell}\hat{n}_{\hat{\ell},q}]$ to $\mathcal{S}_{\ell}$.
    \item {\em Assignment to $\mathcal{S}_\infty$}: Finally, we assign the remaining products $\mathcal{G}_q[\sum_{{\ell} \leq L}\hat{n}_{{\ell},q}+1, |\mathcal{G}_q|]$ to $\mathcal{S}_{\infty}$.
\end{itemize}

\subsection{Analysis: Size, highest-index, and prefix properties}\label{sec_ptas_analysis}

We proceed by arguing that the partition $\mathcal{S} = ({\cal S}_1, \ldots ,{\cal S}_L, {\cal S}_\infty)$, obtained in Sections~\ref{sec_asgn_heavy} and~\ref{sec_asgn_light}, is indeed good, meaning that it satisfies properties~\ref{good_assign_1}-\ref{good_assign_4}. In the upcoming analysis, we focus on the scenario where our guessing procedure for $\{{\beta}^{\uparrow}_{\ell, q}\}_{\ell \in [L], q \in \mathcal{Q}_{\myheavy}}$ is successful, meaning that $\hat{\beta}_{\ell,q} = \beta^{\uparrow}_{\ell,q}$ for every $\ell \in [L]$ and $q \in \mathcal{Q}_{\myheavy}$. At least intuitively, properties~\ref{good_assign_1} and \ref{good_assign_2} are met since each set $\mathcal{S}_{\ell}$ is assigned exactly $\beta^{\uparrow}_{\ell,q}$ products from every weight class $q \in \mathcal{Q}_{\myheavy}$, and at most $\beta^{\uparrow}_{\ell,q}$ products from every weight class $q \in \mathcal{Q}_{\mylight}$. Along these lines, properties~\ref{good_assign_3} and~\ref{good_assign_4} will follow by combining this observation with the assignment of products from each weight class $\mathcal{G}_q$ to each set $\mathcal{S}_{\ell}$ by order of weakly-increasing weights, similarly to the sorted-within-class assignment $\mathcal{A}^{\uparrow}$. Formally, the next claim, whose proof is provided in Appendix~\ref{app_clm_wbq_1}, shows that $\mathcal{S}$ satisfies properties \ref{good_assign_1}, \ref{good_assign_3}, and \ref{good_assign_4} of good partitions, as stated in Section~\ref{sec_good_part}.

\begin{lemma}\label{lem:clm_wbq_1}
    The partition $\mathcal{S} = (\mathcal{S}_1, \ldots, \mathcal{S}_L, \mathcal{S}_\infty)$ satisfies the following properties:
    \begin{itemize}
        \item {\em Bounded size:} $|\mathcal{S}_\ell| \leq \beta^{\uparrow}_\ell$, for every $\ell \in [L]$.         
    
        \item {\em Highest-index weight class:} $\mathcal{S}_\ell \cap \mathcal{G}_{q^{\uparrow}_\ell} \neq \emptyset$, for every $\ell \in [L]$ with  $q^{\uparrow}_\ell \in {\cal Q}_{\myheavy}$. 
    
        \item {\em Prefix subsets:} $\bigcup_{\ell \in [L]} \mathcal{S}_{\ell} \subseteq  \bigcup_{\ell \in [L]_0} \mathcal{A}^{\uparrow}(\mathcal{B}^{\uparrow}_{\ell}) $. 
    \end{itemize}
\end{lemma}

\subsection{Analysis: Weight property} \label{sec_ptas_analysis_cont}

We conclude our analysis by showing that $\mathcal{S} = (\mathcal{S}_{1}, \ldots, \mathcal{S}_L, \mathcal{S}_\infty)$ satisfies property~\ref{good_assign_2}, meaning that $(1-\eps)\cdot{\mathcal{W}}^{\uparrow}_\ell - \eps^4 w_{\max} \leq w(\mathcal{S}_\ell) \leq {\mathcal{W}}^{\uparrow}_\ell$, for every $\ell \in [L]$. To this end, let us decompose the weight of each set $\mathcal{S}_{\ell}$ between heavy and light classes, namely,
\begin{eqnarray} \label{prop_2_break}
    w(\mathcal{S}_{\ell}) & = & \underbrace{\sum_{q \in \mathcal{Q}_{\myheavy}}w(\mathcal{S}_{\ell} \cap \mathcal{G}_q)}_{\mathcal{W}^{\myheavy}_{\ell}}  + \underbrace{\sum_{q \in \mathcal{Q}_{\mylight}}w(\mathcal{S}_{\ell} \cap \mathcal{G}_q)}_{\mathcal{W}^{\mylight}_{\ell}} \ . 
\end{eqnarray} 
 
\paragraph{Exact expression for $\boldsymbol{\mathcal{W}^{\myheavy}_{\ell}}$.}
Starting with the rather straightforward part, the next claim argues that $\mathcal{W}^{\myheavy}_{\ell}$ identifies with the total weight of heavy products within each block $\mathcal{B}^{\uparrow}_{\ell}$ of the sorted-within-class assignment $\mathcal{A}^{\uparrow}$.
\begin{claim}\label{clm_w_heavy}
    $\mathcal{W}^{\myheavy}_{\ell} = \sum_{q \in \mathcal{Q}_{\myheavy}}w(\mathcal{A}^{\uparrow}(\mathcal{B}^{\uparrow}_{\ell}) \cap \mathcal{G}_q)$.
\end{claim}
\begin{proof}
To derive this result, it suffices to show that $w(\mathcal{S}_\ell \cap \mathcal{G}_q) = w( \mathcal{A}^\uparrow(\mathcal{B}^\uparrow_\ell) \cap \mathcal{G}_q )$, for every $q \in \mathcal{Q}_{\myheavy}$. To this end, due to having $\hat{\beta}_{\ell,q} = \beta^{\uparrow}_{\ell,q}$ for every $\ell \in [L]$ and $q \in \mathcal{Q}_{\myheavy}$, each set $\mathcal{S}_{\ell}$ is assigned the set of products $\mathcal{G}_q[\sum_{\hat{\ell} \leq \ell-1}\beta^{\uparrow}_{\hat{\ell},q}+1, \sum_{\hat{\ell}\leq \ell}\beta^{\uparrow}_{\hat{\ell},q}]$
out of ${\cal G}_q$, as explained in Section~\ref{sec_asgn_heavy}. Concurrently, since the assignment $\mathcal{A}^{\uparrow}$ is sorted-within class, the latter set is precisely $\mathcal{A}^{\uparrow}(\mathcal{B}^{\uparrow}_{\ell}) \cap \mathcal{G}_q$, since the block $\mathcal{B}^{\uparrow}_0$ does not contain any heavy products (see Section~\ref{subsec:block_bw}).  As a result, $w(\mathcal{S}_\ell \cap \mathcal{G}_q) = w(\mathcal{A}^{\uparrow}(\mathcal{B}^{\uparrow}_{\ell}) \cap \mathcal{G}_q)$.
\end{proof}

\paragraph{Bounding $\boldsymbol{\mathcal{W}^{\mylight}_{\ell}}$.} We now move on to relate between $\mathcal{W}^{\mylight}_{\ell}$ and the total weight of the light products within block $\mathcal{B}^{\uparrow}_{\ell}$ of the assignment $\mathcal{A}^{\uparrow}$. We begin by lower-bounding $\mathcal{W}^{\mylight}_{\ell}$, showing that this term does not deviate much below $\sum_{q \in \mathcal{Q}_{\mylight}}w(\mathcal{A}^{\uparrow}(\mathcal{B}^{\uparrow}_{\ell}) \cap \mathcal{G}_q)$.

\begin{claim}\label{clm_light_1}
    $\mathcal{W}^{\mylight}_{\ell} \geq(1-\eps) \cdot \sum_{q \in \mathcal{Q}_{\mylight}}  w(\mathcal{A}^{\uparrow}(\mathcal{B}_{\ell}) \cap \mathcal{G}_q) - \eps^4 w_{\max}$.
\end{claim}

\begin{proof}
    As explained in Section~\ref{sec_asgn_light}, when products from each light class $\mathcal{G}_q$ are assigned to $\mathcal{S}_{\ell}$, there are two scenarios to consider. When $\hat{\cal W}_{\ell,q} =0$, no products are assigned, and therefore, $w(\mathcal{S}_{\ell} \cap \mathcal{G}_q) = 0$. When $\hat{\cal W}_{\ell,q} > 0$, the products from $\mathcal{G}_q$ we assign to $\mathcal{S}_\ell$ are guaranteed to have a total weight of at least $(1-\eps)\cdot \hat{\mathcal{W}}_{\ell,q}$. In either case, 
    \begin{eqnarray}
        w(\mathcal{S}_{\ell} \cap \mathcal{G}_q) & \geq & (1-\eps)\cdot \hat{\cal W}_{\ell,q}  \nonumber \\
        & \geq & (1-\eps)\cdot\left(\mathcal{W}^{\uparrow}_{\ell,q}-\frac{\eps^4w_{\max}}{|\mathcal{Q}|}\right)  \label{bound_light_1}\\
        &\geq& (1-\eps)\cdot\mathcal{W}^{\uparrow}_{\ell,q}-\frac{\eps^4w_{\max}}{|\mathcal{Q}|} \nonumber \\
        & = & (1-\eps)\cdot w(\mathcal{A}^{\uparrow}(\mathcal{B}^{\uparrow}_{\ell}) \cap \mathcal{G}_q) -\frac{\eps^4w_{\max}}{|\mathcal{Q}|} \ , \label{bound_light_2} 
    \end{eqnarray}
    where inequality~\eqref{bound_light_1} follows from the relationship between $\hat{\cal W}_{\ell,q}$ and $\mathcal{W}^{\uparrow}_{\ell,q}$, stated in inequality~\eqref{eqn_w_property}.
    By summing inequality~\eqref{bound_light_2} over all light classes, it follows that
    \begin{eqnarray*}
    \mathcal{W}^{\mylight}_\ell
    &=& \sum_{q \in \mathcal{Q}_{\mylight}} w(\mathcal{S}_\ell \cap \mathcal{G}_q) \\
    &\geq& (1 - \eps) \cdot\sum_{q \in \mathcal{Q}_{\mylight}}   w(\mathcal{A}^{\uparrow}(\mathcal{B}^{\uparrow}_\ell) \cap \mathcal{G}_q) - \frac{|{\cal Q}_{\mylight}|}{|{\cal Q}|} \cdot \eps^4 w_{\max}  \\
    & \geq & (1-\eps)\cdot \sum_{q \in \mathcal{Q}_{\mylight}}  w(\mathcal{A}^{\uparrow}(\mathcal{B}^{\uparrow}_{\ell}) \cap \mathcal{G}_q) - \eps^4 w_{\max} \ . 
    \end{eqnarray*}
\end{proof}

Next, we turn our attention to upper-bound $\mathcal{W}^{\mylight}_{\ell}$, showing that this term does not exceed the total weight of light products assigned to the block $\mathcal{B}^{\uparrow}_{\ell}$.

\begin{claim}\label{clm_light_2}
    $\mathcal{W}^{\mylight}_{\ell} \leq \sum_{q \in \mathcal{Q}_{\mylight}} w(\mathcal{A}^{\uparrow}(\mathcal{B}_{\ell}) \cap \mathcal{G}_q)$.
\end{claim}
\begin{proof}
To derive this result, it suffices to show that $w(\mathcal{S}_\ell \cap \mathcal{G}_q) \leq w( \mathcal{A}^\uparrow(\mathcal{B}^\uparrow_\ell) \cap \mathcal{G}_q )$, for every $q \in \mathcal{Q}_{\mylight}$. As explained in Section~\ref{sec_asgn_light}, the set of products from $\mathcal{G}_q$ we assign to $\mathcal{S}_{\ell}$ is exactly $\mathcal{G}_q[\sum_{1\leq\hat{\ell} \leq \ell-1}\hat{n}_{\hat{\ell},q}+1, \sum_{1\leq\hat{\ell}\leq \ell }\hat{n}_{\hat{\ell},q}]$, and therefore,
\begin{eqnarray}        w(\mathcal{S}_{\ell}\cap\mathcal{G}_q) & = & w\left(\mathcal{G}_q\left[\sum_{1\leq\hat{\ell} \leq \ell-1}\hat{n}_{\hat{\ell},q}+1, \sum_{1\leq\hat{\ell}\leq \ell }\hat{n}_{\hat{\ell},q}\right]\right) \nonumber \\
& \leq &  w\left(\mathcal{G}_q\left[\sum_{0\leq\hat{\ell} \leq \ell-1}|\mathcal{A}^{\uparrow}(\mathcal{B}^{\uparrow}_{\hat{\ell}}) \cap \mathcal{G}_q| + 1 , \sum_{0 \leq\hat{\ell}\leq \ell }|\mathcal{A}^{\uparrow}(\mathcal{B}^{\uparrow}_{\hat{\ell}}) \cap \mathcal{G}_q|\right]\right) \label{eqn:clm_light_2_eq1} \\
& = & w(\mathcal{A}^{\uparrow}(\mathcal{B}^{\uparrow}_{\ell}) \cap \mathcal{G}_q) \ . \nonumber
\end{eqnarray}
To better understand where inequality~\eqref{eqn:clm_light_2_eq1} is coming from, we first observe that its left-hand-side is summing the weight of $\hat{n}_{\ell,q}$ consecutive products in $\langle 1, q \rangle, \ldots, \langle |\mathcal{G}_q|, q \rangle$, which is an ordering of $\mathcal{G}_q$ by weakly-increasing weight. In contrast, the right-hand-side of~\eqref{eqn:clm_light_2_eq1} is summing over $|\mathcal{A}^{\uparrow}(\mathcal{B}^{\uparrow}_{\ell}) \cap \mathcal{G}_q|$ consecutive products. As explained in Section~\ref{sec_asgn_light}, we have $\hat{n}_{\ell,q} \leq |\mathcal{A}^{\uparrow}(\mathcal{B}^{\uparrow}_{\ell}) \cap \mathcal{G}_q|$ for every $\ell \in [L]$, implying that the left-hand-side is summing over fewer terms, starting at an earlier point in the (weakly-increasing) sequence, which is why it is upper-bounded by the right-hand-side.
\end{proof}

\paragraph{Putting everything together.} Finally, by decomposing $w(\mathcal{S}_{\ell})$ into $\mathcal{W}^{\myheavy}_{\ell}$ and $\mathcal{W}^{\mylight}_{\ell}$, as in \eqref{prop_2_break}, and bounding each of these terms, we prove that $\mathcal{S}= (\mathcal{S}_1, \ldots, \mathcal{S}_L, \mathcal{S}_{\infty})$ indeed satisfies property~\ref{good_assign_2}, arguing that $(1-\eps)\cdot{\mathcal{W}}^{\uparrow}_\ell - \eps^4 w_{\max} \leq w(\mathcal{S}_\ell) \leq {\mathcal{W}}^{\uparrow}_\ell$, for every $\ell \in [L]$. 
Starting with the lower bound, we observe that
\begin{eqnarray*}
    w(\mathcal{S}_\ell) & = & \mathcal{W}^{\myheavy}_{\ell} + \mathcal{W}^{\mylight}_{\ell}  \\
    &\geq &  \sum_{q \in \mathcal{Q}_{\myheavy}}  w(\mathcal{A}^{\uparrow}(\mathcal{B}^{\uparrow}_{\ell}) \cap \mathcal{G}_q) + (1-\eps)\cdot\sum_{q \in \mathcal{Q}_{\mylight}}  w(\mathcal{A}^{\uparrow}(\mathcal{B}^{\uparrow}_{\ell}) \cap \mathcal{G}_q)- \eps^4 w_{\max}  \\
    & \geq & (1-\eps)\cdot\sum_{q \in [Q]}  w(\mathcal{A}^{\uparrow}(\mathcal{B}^{\uparrow}_{\ell}) \cap \mathcal{G}_q)- \eps^4 w_{\max} \\
    & = & (1-\eps)\cdot{\mathcal{W}}^{\uparrow}_\ell - \eps^4 w_{\max} \ , 
\end{eqnarray*}
where the first inequality above follows from Claims~\ref{clm_w_heavy} and \ref{clm_light_1}. Now, to upper-bound $w(\mathcal{S}_{\ell})$, note that
\begin{eqnarray*}
    w(\mathcal{S}_\ell) & = & \mathcal{W}^{\myheavy}_{\ell} + \mathcal{W}^{\mylight}_{\ell}  \\
    &\leq & \sum_{q \in \mathcal{Q}_{\myheavy}}  w(\mathcal{A}^{\uparrow}(\mathcal{B}^{\uparrow}_{\ell}) \cap \mathcal{G}_q) + \sum_{q \in \mathcal{Q}_{\mylight}}  w(\mathcal{A}^{\uparrow}(\mathcal{B}^{\uparrow}_{\ell}) \cap \mathcal{G}_q) \\
    & = & {\mathcal{W}}^{\uparrow}_\ell \ ,  
\end{eqnarray*}
where the inequality above is obtained by combining Claims \ref{clm_w_heavy} and \ref{clm_light_2}.

%% file: TEX-Conclusions.tex
\section{Concluding Remarks}

We conclude our work by presenting a number of particularly challenging questions for future research. These prospective avenues take aim at fundamental open questions around fine-grained implementations, asking whether the running time of our approach could be enhanced via new algorithmic ideas. In addition, while our work focuses on market share optimization, we bring attention to the generalized setting of revenue maximization, explaining why it appears to be very difficult to deal with.

\paragraph{EPTAS for market share ranking?} As stated in Theorem~\ref{thm:main_result}, our main algorithmic contribution consists in designing a polynomial-time approximation scheme (PTAS) for the market share ranking problem, admitting an $O(n^{O(\frac{1} {\eps^{6}}\log\frac{1}{\eps})})$-time implementation. That said, one could still wonder about the plausibility of an EPTAS, which is an approximation scheme whose running time can be expressed as $f(\frac{1}{\eps})\cdot n^{O(1)}$, for some arbitrary function $f$ depending only on $\frac{1}{\eps}$. Toward this objective, we remind the reader that, among other ideas, our algorithm employs two enumeration-based guessing procedures, one for the total number $\hat{\beta}_{\ell,q}$ of heavy products within each block, and another for the analogous cumulative weight $\hat{\cal W}_{\ell,q}$ of light products. Currently, these procedures serve as our main bottlenecks toward developing an EPTAS, due to incurring running times of $O({n}^{O(\frac{1}{\eps^2}\log^2(\frac{1}{\eps}))})$ and $O(n^{O({\frac{1}{\eps^{6}}\log\frac{1}{\eps})}})$, respectively. It would be interesting to target these procedures, attempting to prove that more coarse guesses are sufficient to identify near-optimal assignments. 

On the other hand, it is quite possible that an EPTAS simply does not exist. A potential avenue to prove a hardness result of this nature may be inspired by an approach similar to that of \cite{KULIK2010707}, showing that two-dimensional knapsack does not admit an EPTAS. Since market share ranking exhibits certain knapsack-like properties, one might hope to uncover intractability-related connections between these two settings. Additional examples of problems that admit a PTAS but not an EPTAS have previously been discovered by \cite{marx2005efficient}, \cite{HuangChen2006}, \cite{cygan2016closest},  \cite{abboud2022search}, and \cite{tight2024budgeted}.

\paragraph{Efficient dynamic programming approach?} In Section~\ref{subsec:qptas}, we provide a quasi-polynomial-time approximation scheme for the market share ranking problem, admitting an $O ( K^2 n^{ O(\frac{1}{\eps}\log\frac{n}{\eps}) } )$-time implementation. This algorithm relies on exploiting newly-revealed structural properties within a dynamic programming approach, originally proposed by \cite{DerakhshanGMM22}. That said, we still do not know whether further technical insights can be developed to arrive at a true PTAS by these means. Along these lines, an open question is whether various dynamic programming speed-up strategies, such as those of \cite{SegevShaposhnik21} and \cite{RiegerSegev22}, could be leveraged to address this challenge.

\paragraph{Revenue maximization?} Our work focuses on the market share ranking problem, where a retailer wishes to determine a position-to-product assignment ${\cal A} : [n] \to [n]$ whose expected market share $M( {\cal A} )$ is maximized. This expectation is taken with respect to the segment proportions $\theta_1, \ldots, \theta_K$, meaning that $M( {\cal A} ) = \sum_{k \in [K]} \theta_k \cdot M_k( {\cal A} )$, where each market share component is given by $M_k( {\cal A} ) = \frac{ w( C_k^{\cal A} ) }{ 1 + w( C_k^{\cal A} )}$. At present time, the big unknown is general revenue maximization, where each product $i \in [n]$ is associated with a selling price of $r_i$, and our goal is to maximize expected revenue. This measure is specified by $R( {\cal A} ) = \sum_{k \in [K]} \theta_k \cdot R_k( {\cal A}) $, where each component has an individual revenue of $R_k(\mathcal{A}) = \sum_{i \in C^{\mathcal{A}}_k} r_i \cdot\frac{ w_i}{1 + w(C^{\mathcal{A}}_k)}$.

In this formulation, preference weights and prices may be uncorrelated, implying that expensive products could be associated with low preference weights and vice versa, creating a highly unstructured objective function. For instance, while the single-segment market share term $M_k(\mathcal{A}) = \frac{w(C_k^{\mathcal{A}})}{1 + w(C_k^{\mathcal{A}})}$ is monotone and concave in the aggregate weight $w(C_k^{\mathcal{A}})$, elementary examples demonstrate that the revenue term $R_k(\mathcal{A}) = \sum_{i \in C_k^{\mathcal{A}}} r_i \cdot \frac{w_i}{1 + w(C_k^{\mathcal{A}})}$ lacks this basic structure, due to the presence of selling prices. In particular, augmenting any consideration set with an additional product can have opposing effects: While collecting the product's own revenue contribution $r_i \cdot \frac{w_i}{1 + w(C_k^{\mathcal{A}})}$, we simultaneously decrease the choice probabilities, and thus revenue contributions, of all other products. As such, the net effect depends on a complex interplay between product prices and weights, deviating from the cleaner structure of market share ranking. This phenomenon renders natural heuristics as well as more complex approaches ineffective, even on seemingly simple instances. An interesting open question is whether any non-trivial approximation guarantees can be attained for this setting, or whether stronger hardness results can be established.

%% file: TEX-Appendix.tex
%%%%%%%%%%%%%%%%%%%%%%%%%%%%%%%%%%%%
\section{Additional Proofs from Section~\ref{sec:overview}}

\subsection{Proof of Lemma~\ref{lem:market_share_early}} \label{app:proof_tearly} 

By definition of $\Tearly$, the stopping point $s^{\mathcal{A}^{\uparrow}}_k$ of each customer $k$ in this class with respect to the assignment $\mathcal{A}^{\uparrow}$ resides within the block $\mathcal{B}^{\uparrow}_0$. As explained in Section~\ref{subsec:block_bw}, this block is defined as   $\mathcal{B}^{\uparrow}_0 = [1,p_0]$, where $p_0$ is the maximal position $p \in [n]$ for which $w(\mathcal{A}^{\uparrow}[1,p]) < \eps^3 w_{\max}$.  Thus, for the consideration set $C^{\mathcal{A}^{\uparrow}}_k$ of this customer, we have $w(C^{\mathcal{A}^{\uparrow}}_k) < \eps^3 w_{\max}$, and therefore,
\[ M_k ( {\cal A}^{ \uparrow }  ) ~~=~~ \frac{w(C^{\mathcal{A}^{\uparrow}}_k)}{1 + w(C^{\mathcal{A}^{\uparrow}}_k)} ~~\leq~~ \frac{\eps^3 w_{\max}}{1 + \eps^3 w_{\max}} ~~\leq~~ \eps^3 w_{\max} \ . \]

\subsection{Proof of Lemma~\ref{lem:market_share_mid}} \label{app:proof_tmid}

Let us focus on a single midway stopper $k$, namely, one whose stopping point $s^{\mathcal{A}^{\uparrow}}_k$ with respect to the assignment $\mathcal{A}^{\uparrow}$ resides within ${\cal B}^{\uparrow}_{\ell_k}$, for some $\ell_k \in [L]$. Our analysis is based on arguing that, with respect to the assignment $\tilde{\cal A}$, this customer necessarily stops within one of the blocks $\tilde{\cal B}_{\ell_k}, \ldots, \tilde{\cal B}_{L}, \tilde{\cal B}_{\infty}$, as stated in Claim~\ref{clm_tmid_stop} below. Building on this result, our second observation will show that the consideration set $C^{\tilde{\cal A}}_k$ of customer $k$ nearly matches $C^{{\cal A}^{\uparrow}}_k$ in terms of total weight. The proofs of these two claims are presented in Appendices~\ref{proof_a3} and \ref{proof_a4}. 

\begin{claim}\label{clm_tmid_stop}
    $s^{\tilde{\cal A}}_k \in \tilde{\cal B}_{[\ell_k, \infty]}$.
\end{claim}
\begin{claim}\label{clm_tmid_set}
    $w(C^{\tilde{\cal A}}_k) \geq (1-2\eps)\cdot w(C^{{\cal A}^{\uparrow}}_k)-4 \eps^2  w_{\max}$.
\end{claim}

Given Claim~\ref{clm_tmid_set}, we relate between the expected market shares $M_k(\tilde{\cal A})$ and $M_k({\cal A}^{\uparrow})$ of this customer with respect to $\tilde{\cal A}$ and $\mathcal{A}^{\uparrow}$ by observing that
\begin{eqnarray}
    M_k(\tilde{\cal A}) & = &\nonumber \frac{w(C^{\tilde{\cal A}}_k)}{1+w(C^{\tilde{\cal A}}_k)} \\\nonumber\\
    & \geq & \nonumber \frac{(1-2\eps)\cdot w(C^{{\cal A}^{\uparrow}}_k)-4 \eps^2  w_{\max}}{1 + (1-2\eps)\cdot w(C^{{\cal A}^{\uparrow}}_k)-4 \eps^2  w_{\max}} \\\nonumber\\
    & \geq & \nonumber \frac{(1-2\eps)\cdot w(C^{{\cal A}^{\uparrow}}_k)-4 \eps ^2  w_{\max}}{1 + w(C^{{\cal A}^{\uparrow}}_k)} \\\nonumber\\
    & \geq & \nonumber (1-2\eps) \cdot M_k({\cal A}^{\uparrow}) -4 \eps^2   w_{\max} \ .
\end{eqnarray}

\subsection{Proof of Claim~\ref{clm_tmid_stop}}\label{proof_a3} 

As explained in Section~\ref{subsec:convert_b_w}, due to the specific way we converted $\mathcal{S} = ({\cal S}_1, \ldots {\cal S}_L, {\cal S}_{\infty})$ into the assignment $\tilde{\cal A}$, each of our resulting blocks $\tilde{\cal B}_{\ell}$ contains the set of products ${\cal S}_\ell$ and nothing more, for every $\ell \in [L]$, implying that $w( \tilde{\cal A}( \tilde{\cal B}_{\ell} ) ) = w( {\cal S}_\ell )$. Since ${\cal S}$ is a good partition, by property~\ref{good_assign_2} we know that 
\begin{equation}\label{eq_mid_prop}
         w( \tilde{\cal A}( \tilde{\cal B}_{\ell} ) ) ~~\leq~~ {\mathcal{W}}^{\uparrow}_\ell \ . 
    \end{equation}
 Next, for every $\ell \in [L]$ with $\tilde{\cal B}_{\ell} \neq \emptyset$, let us make use of $\tilde{p}_{\ell}$ and ${p}^{\uparrow}_{\ell}$ to designate the highest-index positions of the blocks $\tilde{\cal B}_{\ell}$ and ${\cal B}^{\uparrow}_{\ell}$, respectively, noting that since $|\tilde{\cal B}_{\ell}|=|\mathcal{S}_{\ell}|\leq \beta^{\uparrow}_{\ell} = |\mathcal{B}^{\uparrow}_{\ell}|$ by property~\ref{good_assign_1}, the block $\mathcal{B}^{\uparrow}_{\ell}$ is non-empty as well. With this notation, we observe that
\begin{eqnarray}
    w(\tilde{\cal A}[1, \tilde{p}_{\ell}]) 
    & = & \sum^{\ell}_{\hat{\ell}=1}w(\tilde{\cal A}(\tilde{\cal B}_{\hat{\ell}})) \nonumber\\
    & \leq & \sum^{\ell}_{\hat{\ell}=0}\mathcal{W}^{\uparrow}_{\hat{\ell}} \nonumber\\
    & = & w({\cal A}^{\uparrow}[1, {p}^{\uparrow}_{\ell}]) \ ,  \label{eq_stop_mid_2}
\end{eqnarray}
where  the inequality above is obtained by plugging inequality~\eqref{eq_mid_prop} and adding the extra term $\mathcal{W}^{\uparrow}_0$.

We are now ready to show that $s^{\mathcal{A}^{\uparrow}}_k \in \tilde{\cal B}_{[\ell_k, \infty]}$. For this purpose, it suffices to argue that the stopping condition of customer $k$ is not met at position $\tilde{p}_{\ell^{-}_k}$, where 
$\ell^{-}_k$ is the highest index out of $1, \ldots, \ell_{k}-1$ for which block ${\cal B}_{\ell^{-}_k}$ is non-empty. To this end, we observe that
\begin{eqnarray}
    w(\tilde{\cal A}[1, \tilde{p}_{\ell^{-}_k}]) 
    & \leq & w({\cal A}^{\uparrow}[1, {p}^{\uparrow}_{\ell^{-}_k}]) \label{eq_stop_2}\\
    & < &  r^{k}_{{p}^{\uparrow}_{\ell^{-}_k}} \label{eq_stop_3}\\
    & \leq & \label{eq_stop_4} r^{k}_{\tilde{p}_{\ell^{-}_k}} \ .
\end{eqnarray}
Here, inequality~\eqref{eq_stop_2} is precisely inequality~\eqref{eq_stop_mid_2}, instantiated with $\ell = \ell^{-}_k$. Inequality~\eqref{eq_stop_3} holds since $s^{\mathcal{A}^{\uparrow}}_k \in \mathcal{B}^{\uparrow}_{{\ell}_k}$ by our initial assumption, implying that the stopping condition of this customer is not met at position $p^{\uparrow}_{\ell^{-}_k}$. Finally, we arrive at inequality~\eqref{eq_stop_4} by recalling that the sequence $r^k_1, \ldots, r^k_n$ is weakly decreasing, implying that we have $r^{k}_{\tilde{p}_{\ell^{-}_k}} \geq r^{k}_{{p}^{\uparrow}_{\ell^{-}_k}}$, since
\begin{equation}\label{stop_pp}
    \tilde{p}_{\ell^{-}_k} ~~=~~ \sum^{\ell^{-}_k}_{\ell=1} |\mathcal{S}_{\ell}| ~~\leq~~ \sum^{\ell^{-}_k}_{\ell=1} \beta^{\uparrow}_{\ell} ~~\leq~~ \sum^{\ell^{-}_k}_{\ell=0}|\mathcal{B}^{\uparrow}_{\ell}| ~~=~~ p^{\uparrow}_{\ell^{-}_k} \ .
\end{equation}

\subsection{Proof of Claim~\ref{clm_tmid_set}} \label{proof_a4}
By Claim~\ref{clm_tmid_stop}, we know that the consideration set $C^{\tilde{\cal A}}_k$ of customer $k$ consists of all products in $\tilde{\cal B}_{[1, \ell_k -1]}$, along with at least one product from $\tilde{\cal B}_{\ell_k}$. We denote the first position of the latter block by $\tilde{\eta}_{\ell_k}$; similarly, ${\eta}^{\uparrow}_{\ell_k}$ will denote the first position of $\mathcal{B}^{\uparrow}_{\ell_k}$. To relate between $w(C^{\tilde{\cal A}}_k)$ and  $w(C^{{\cal A}^{\uparrow}}_k)$, we first compare the cumulative weight of the products assigned to $\tilde{\cal B}_{[1, \ell_k -1]}$ and ${\cal B}^{\uparrow}_{[1, \ell_k -1]}$ by $\tilde{\cal A}$ and $\mathcal{A}^{\uparrow}$, respectively, and then compare between the products assigned to positions $\tilde{\eta}_{\ell_k}$ and ${\eta}^{\uparrow}_{\ell_k}$. Both claims are proven at the end of this section.

\begin{claim}\label{clm_tmid_first_component}
    $w(\tilde{\cal A}(\tilde{\cal B}_{[1,\ell_k-1]})) \geq (1-\eps) \cdot \sum^{\ell_k-1}_{\ell=1} \mathcal{W}^{\uparrow}_\ell - 2 \eps^2  w_{\max} \ .$
\end{claim}

\begin{claim}\label{clm_tmid_second_component}
    $ w(\tilde{\cal A}(\tilde{\eta}_{\ell_k})) \geq  (1-\eps) \cdot w({\cal A}^{\uparrow}(\eta^{\uparrow}_{\ell_k}))-\eps^2 w_{\max} \ .$
\end{claim}

Putting both bounds together, we conclude that
\begin{eqnarray}
    w(C^{\tilde{\cal A}}_k) & \geq & w(\tilde{\cal A}(\tilde{\cal B}_{[1, \ell_k-1]})) + w(\tilde{\cal A}(\tilde{\eta}_{\ell_k}))   \nonumber\\ 
    &\geq& \nonumber (1-\eps) \cdot \sum^{\ell_k-1}_{\ell=1} \mathcal{W}^{\uparrow}_\ell - 2 \eps^2  w_{\max} + (1-\eps) \cdot w({\cal A}^{\uparrow}(\eta^{\uparrow}_{\ell_k}))-\eps^2 w_{\max} \\
    & = & \nonumber (1-\eps) \cdot \left(\sum^{\ell_k-1}_{\ell=0} \mathcal{W}^{\uparrow}_\ell-\mathcal{W}^{\uparrow}_0\right) + (1-\eps) \cdot w({\cal A}^{\uparrow}(\eta^{\uparrow}_{\ell_k})) - 3 \eps^2  w_{\max}\\
    &\geq & \label{tmid_ineq_1} (1-\eps) \cdot \sum^{\ell_k-1}_{\ell=0} \mathcal{W}^{\uparrow}_\ell + (1-\eps) \cdot w({\cal A}^{\uparrow}(\eta^{\uparrow}_{\ell_k})) - 4 \eps^2  w_{\max}\\
    & \geq & \label{tmid_ineq_2} (1-\eps)(1+\eps)^{\ell_k-1} \cdot \eps^3 w_{\max}- 4 \eps^2  w_{\max}\\
    & \geq & \label{tmid_ineq_3} (1-\eps)^2\cdot w(C^{{\cal A}^{\uparrow}}_k)- 4 \eps^2  w_{\max} \\
    & \geq & \nonumber (1-2\eps)\cdot w(C^{{\cal A}^{\uparrow}}_k)-4 \eps^2  w_{\max} \ . 
\end{eqnarray}
Here, inequality~\eqref{tmid_ineq_1} holds since $\mathcal{W}^{\uparrow}_0 = w(\mathcal{A}^{\uparrow}(\mathcal{B}^{\uparrow}_0)) < \eps^3 w_{\max}$, by definition of $\mathcal{B}^{\uparrow}_0$.
To understand where inequality~\eqref{tmid_ineq_2} is coming from, note that 
\[\sum^{\ell_k-1}_{\ell=0} \mathcal{W}^{\uparrow}_{\ell} + w({\cal A}^{\uparrow}(\eta^{\uparrow}_{\ell_k})) ~~=~~ w({\cal A}^{\uparrow}({\cal B}^{\uparrow}_{[0, \ell_k -1]})) + w({\cal A}^{\uparrow}(\eta^{\uparrow}_{\ell_k}))  ~~\geq~~ (1+\eps)^{\ell_k-1} \cdot \eps^3 w_{\max} \ , \]
according to the definition of $\mathcal{B}^{\uparrow}_{\ell_k - 1}$ in Section~\ref{subsec:block_bw}. Finally, inequality~\eqref{tmid_ineq_3} follows by recalling that $s^{\mathcal{A}^{\uparrow}}_k \in {\cal B}^{\uparrow}_{\ell_k}$, and therefore, 
\[w(C^{\mathcal{A}^{\uparrow}}_k) ~~\leq~~ w({\cal A}^{\uparrow}({\cal B}^{\uparrow}_{[0, \ell_k ]})) ~~\leq~~  (1+\eps)^{\ell_k}\cdot \eps^3 w_{\max} \ .\]

\paragraph{Proof of Claim~\ref{clm_tmid_first_component}.} To derive the desired bound, we observe that
\begin{eqnarray} 
w(\tilde{\cal A}(\tilde{\cal B}_{[1, \ell_k -1]})) & = & \nonumber  \sum^{\ell_k-1}_{\ell=1}w({\cal S}_{\ell})   \\ 
& \geq & \label{tmid_1}  \sum^{\ell_k-1}_{\ell=1} \left( (1-\eps) \cdot{\mathcal{W}}^{\uparrow}_\ell - \eps^4 w_{\max} \right)   \\ \nonumber
& \geq & \label{tmid_2} (1-\eps) \cdot \sum^{\ell_k-1}_{\ell=1} \mathcal{W}^{\uparrow}_\ell - L \eps^4 w_{\max} \\
& \geq & \label{tmid_3} (1-\eps) \cdot \sum^{\ell_k-1}_{\ell=1} \mathcal{W}^{\uparrow}_\ell - 2 \eps^2  w_{\max} \ .
\end{eqnarray} 
Here, inequality~\eqref{tmid_1} holds due to property~\ref{good_assign_2} of good partitions, stating in particular that  $w({\cal S}_{\ell}) \geq (1-\eps)\cdot {\mathcal{W}}^{\uparrow}_\ell - \eps^4 w_{\max}$. Inequality~\eqref{tmid_3} is obtained by noting that $ L  =   \lceil \log_{1+\eps} (\frac{1}{\eps^{4}}) \rceil  \leq   \frac{4\ln(\frac{1}{\eps})}{\ln(1+\eps)} + 1 \leq \frac{2}{\eps^2}$, where the last transition follows from elementary calculus arguments. 

\paragraph{Proof of Claim~\ref{clm_tmid_second_component}.} To upper-bound the weight of product $\tilde{\cal A}(\tilde{\eta}_{\ell_k})$, let us recall that $q^{\uparrow}_{\ell_k}$ stands for the maximal index $q$ for which $\mathcal{B}^{\uparrow}_{\ell_k} \cap \mathcal{G}_q \neq \emptyset$. We proceed by considering two possible scenarios, depending on whether $q^{\uparrow}_{\ell_k}$ belongs to $\mathcal{Q}_{\myheavy}$ or $\mathcal{Q}_{\mylight}$. 
\begin{itemize}
    \item {\em When $q^{\uparrow}_{\ell_k} \in \mathcal{Q}_\myheavy$}: In this case, due to the way we construct $\tilde{\cal A}$, specifically by placing a product from class $\mathcal{G}_{{q}^{\uparrow}_{\ell_k}}$ in position $\tilde{\eta}_{\ell_k}$ (see Section~\ref{subsec:convert_b_w}), we know that $w(\tilde{\cal A}(\tilde{\eta}_{\ell_k})) \geq (1+\eps)^{q^{\uparrow}_{\ell_k}-1} \cdot \frac{\eps^2}{2n} \cdot w_{\max}$.
    
    \item {\em When $q^{\uparrow}_{\ell_k} \in \mathcal{Q}_\mylight$}: Here, regardless of product $\tilde{\cal A}(\tilde{\eta}_{\ell_k})$, every product in $\mathcal{A}^{\uparrow}(\mathcal{B}^{\uparrow}_{\ell_k})$ has a weight of at most $\eps^2  w_{\max}$.
\end{itemize}
Therefore,
\begin{eqnarray}
    w(\tilde{\cal A}(\tilde{\eta}_{\ell_k})) & \geq & \nonumber
 \min \left\{(1+\eps)^{q^{\uparrow}_{\ell_k}-1} \cdot \frac{\eps^2}{2n} \cdot  w_{\max}, w({\cal A}^{\uparrow}({\eta}^{\uparrow}_{\ell_k}))-\eps^2 w_{\max}\right\} \\\nonumber\\
 & \geq & \label{tmid_4}
 \min \left\{(1-\eps) \cdot w({\cal A}^{\uparrow}({\eta}^{\uparrow}_{\ell_k})), w({\cal A}^{\uparrow}({\eta}^{\uparrow}_{\ell_k}))-\eps^2 w_{\max}\right\} \\\nonumber\\
 & \geq & \nonumber (1-\eps) \cdot w({\cal A}^{\uparrow}(\eta^{\uparrow}_{\ell_k}))-\eps^2 w_{\max} \ , 
\end{eqnarray}
where inequality~\eqref{tmid_4} follows by recalling that the highest-index weight class appearing in block $\mathcal{ B}^{\uparrow}_{\ell_k}$ is $q^{\uparrow}_{\ell_k}$, and therefore $ w({\cal A}^{\uparrow}({\eta}^{\uparrow}_{\ell_k})) \leq (1+\eps)^{q^{\uparrow}_{\ell_k}} \cdot \frac{\eps^2}{2n} \cdot w_{\max}$.

\subsection{Proof of Lemma~\ref{lem:market_share_late}} \label{app:proof_tlate}
Let us focus on a single late stopper $k$, whose stopping point $s^{{\cal A}^{\uparrow}}_k$ resides within block $\mathcal{B}^{\uparrow}_{\infty}$. For simplicity of notation, we assume that block $\tilde{\cal B}_L$ is non-empty; the opposite case can be handled via nearly-identical arguments. Recalling that $\tilde{p}_L$ and $p^{\uparrow}_L$ denote the highest-index positions of blocks $\tilde{\cal B}_L$ and $\mathcal{B}^{\uparrow}_L$, respectively, it is easy to verify that $w(\tilde{\cal A}[1, \tilde{p}_L]) \leq r^k_{\tilde{p}_L}$, simply by duplicating the proof of inequality~\eqref{eq_stop_4} for midway stoppers. As such, the stopping point $s^{\tilde{\cal A}}_k$ resides within $\tilde{\cal B}_{\infty}$, and to connect between the consideration sets $C^{\tilde{\mathcal{A}}}_k$ and $C^{\mathcal{A}^{\uparrow}}_k$, we consider two possible scenarios, depending on the relation between $s^{\tilde{\cal A}}_k$ and $s^{{\cal A}^{\uparrow}}_k$. 

\paragraph{Case 1: $\boldsymbol{s^{\tilde{\cal A}}_k \geq s^{{\cal A}^{\uparrow}}_k}$.} Let us begin by observing that $w(\tilde{\cal A}(\tilde{\cal B}_{[1,L]}))$ and $w({\cal A}^{\uparrow}({\cal B}^{\uparrow}_{[0,L]}))$ can be related by duplicating the proof of Claim~\ref{clm_tmid_set}, with respect to a customer whose stopping point resides within block $\tilde{ \cal B}_L$, to obtain
\begin{eqnarray}
    w(\tilde{\cal A}(\tilde{\cal B}_{[1,L]}))
    & \geq &  (1-2\eps) \cdot w({\cal A}^{\uparrow}({\cal B}^{\uparrow}_{[0,L]}))- 4\eps^2 w_{\max} \label{late_stop_1} \ .
\end{eqnarray}
Next, we proceed by relating between $w(\tilde{\cal A}[\tilde{p}_L + 1, s^{\tilde{\cal A}}_k] )$ and $w({\cal A}^{\uparrow}[p^{\uparrow}_L + 1, s^{{\cal A}^{\uparrow}}_k])$. First, by substituting $\ell^-_k$ with $L$ in inequality~\eqref{stop_pp}, we infer that $|\tilde{\cal B}_{[1,L]}|\leq |\mathcal{B}^{\uparrow}_{[0,L]}|$. As such, combined with our case hypothesis that $s^{\tilde{\cal A}}_k \geq s^{{\cal A}^{\uparrow}}_k$, it follows that $|\tilde{\cal A}[\tilde{p}_L + 1, s^{\tilde{\cal A}}_k] | \geq |{\cal A}^{\uparrow}[p^{\uparrow}_L + 1, s^{{\cal A}^{\uparrow}}_k]|$, meaning that when comparing the consideration set of customer $k$ with respect to both assignments, the one formed by $\tilde{\cal A}$ contains more items from block $\tilde{\cal B}_{\infty}$ than the one formed by $\mathcal{A}^{\uparrow}$ from block ${\cal B}^{\uparrow}_{\infty}$. In turn, property~\ref{good_assign_4} of good partitions states that the products appearing in $\mathcal{S}_1, \ldots, \mathcal{S}_L$ are a subset of ${\cal A}^{\uparrow}({\cal B}^{\uparrow}_{[0,L]})$, implying that $\tilde{\cal A}(\tilde{\cal B}_{[1,L]}) \subseteq {\cal A}^{\uparrow}({\cal B}^{\uparrow}_{[0,L]})$, and therefore, ${\cal A}^{\uparrow}(\mathcal{B^{\uparrow}_{\infty}})\subseteq \tilde{\cal A}(\tilde{\cal B}_{\infty})$.
Consequently, since we ensure that the products in $\tilde{\cal B}_{\infty}$ are sorted by order of weakly-decreasing weight, 
\begin{eqnarray}
    w(\tilde{\cal A}[\tilde{p}_L + 1, s^{\tilde{\cal A}}_k] ) ~~\geq~~ w({\cal A}^{\uparrow}[p^{\uparrow}_L + 1, s^{{\cal A}^{\uparrow}}_k]) \ .  \label{late_stop_2}
\end{eqnarray}
Putting \eqref{late_stop_1} and \eqref{late_stop_2} together, we have
\begin{eqnarray}
    w(C^{\tilde{\cal A}}_k) &=& w(\tilde{\cal A}(\tilde{\cal B}_{[1,L]})) + w(\tilde{\cal A}[\tilde{p}_L + 1, s^{\tilde{\cal A}}_k] ) \nonumber \\
    & \geq & (1-2\eps) \cdot w({\cal A}^{\uparrow}({\cal B}^{\uparrow}_{[0,L]}))- 4\eps^2 w_{\max} +  w({\cal A}^{\uparrow}[p^{\uparrow}_L + 1, s^{{\cal A}^{\uparrow}}_k]) \nonumber \\
    & \geq & (1-2\eps)\cdot w(C^{\mathcal{A}^{\uparrow}}_k)-4\eps^2 w_{\max} \ \nonumber .
\end{eqnarray}

\paragraph{Case 2: $\boldsymbol{s^{\tilde{\cal A}}_k < s^{{\cal A}^{\uparrow}}_k}$.} Recalling that the stopping point $s^{\tilde{\cal A}}_k$ corresponds to the earliest position $p$ for which $w(\tilde{\cal A}[1,p]) \geq r^k_{p}$, we must have
\begin{eqnarray}
    w(C^{\tilde{\cal A}}_k) &=& w(\tilde{\cal A}[1, s^{\tilde{\cal A}}_k]) \nonumber \\
    & \geq & r^k_{s^{\tilde{\cal A}}_k} \nonumber\\
    & \geq & \label{tlate_0}  r^k_{s^{{\cal A}^{\uparrow}}_k-1} \\
    & >  & \label{tlate_1} w({\cal A}^{\uparrow}[1, s^{{\cal A}^{\uparrow}}_k-1]) \\
    & \geq & \nonumber w({\cal A}^{\uparrow}[1, s^{{\cal A}^{\uparrow}}_k]) - w_{\max} \\
    & \geq & \label{tlate_2} (1-\eps)\cdot w(C^{{\cal A}^{\uparrow}}_k) \ .
\end{eqnarray}
Here, inequality~\eqref{tlate_0} holds since $r^k_1 \geq \cdots\geq r^k_n$ and since $s^{\tilde{\cal A}}_k < s^{{\cal A}^{\uparrow}}_k$, by the case hypothesis.
Inequality~\eqref{tlate_1} follows by noting that $s^{{\cal A}^{\uparrow}}_k$ is the earliest position $p$ with $w({\cal A}^{\uparrow}[1,p]) \geq r^k_{p}$, and therefore, $w({\cal A}^{\uparrow}[1, s^{{\cal A}^{\uparrow}}_k-1]) < r^k_{s^{{\cal A}^{\uparrow}}_k-1}$. Finally, inequality~\eqref{tlate_2} holds since customer $k$ is a late stopper, meaning that $w(C^{{\cal A}^{\uparrow}}_k) \geq w({\cal A}^{\uparrow}[1, p_L +1]) \geq \frac{w_{\max}}{\eps}$, since for any position $p > p^{\uparrow}_L$, we hit a cumulative weight of $w(\mathcal{A}^{\uparrow}[1,p]) \geq (1+\eps)^L\cdot\eps^3 w_{\max} \geq \frac{w_{\max}}{\eps}$.

\paragraph{Summary.} With respect to the assignment $\tilde{\cal A}$, we conclude that customer $k$ has an expected market share of
\begin{eqnarray}
    M_k(\tilde{\mathcal{A}}) & = & \nonumber \frac{w(C^{\tilde{\mathcal{A}}}_k)}{1 + w(C^{\tilde{\mathcal{A}}}_k)} \\\nonumber\\ & \geq & \nonumber \frac{(1-2\eps)\cdot w(C^{\mathcal{A}^{\uparrow}}_k)-4\eps^2 w_{\max}}{1 + (1-2\eps)\cdot w(C^{\mathcal{A}^{\uparrow}}_k)-4\eps^2 w_{\max}} \\\nonumber\\
    & \geq & \nonumber \frac{(1-2\eps)\cdot w(C^{\mathcal{A}^{\uparrow}}_k)-4 \eps^2 w_{\max}}{1 + w(C^{\mathcal{A}^{\uparrow}}_k)} \\\nonumber\\
    & \geq & \nonumber (1-2\eps)\cdot M_k(\mathcal{A}^{\uparrow})-4 \eps^2 w_{\max} \ . 
\end{eqnarray}

%%%%%%%%%%%%%%%%%%%%%%%%%%%%%%%%%%%%
\section{Additional Proofs from Section~\ref{sec:compute_partition}}

\subsection{Estimating light-class assignment weights}\label{app_clm_guess_w}

In what follows, we describe how the family of estimates $\mathcal{F}_W$ is constructed, ensuring that it contains at least one set of guesses $\{\hat{\cal W}_{\ell,q}\}_{\ell \in [L], q \in \mathcal{Q}_{\mylight}}$ satisfying $\mathcal{W}^{\uparrow}_{\ell, q} -\frac{\eps^4 w_{\max}}{|\mathcal{Q}|} \leq \hat{\cal W}_{\ell, q} \leq \mathcal{W}^{\uparrow}_{\ell, q}$, for every $\ell \in [L]$ and $q \in \mathcal{Q}_{\mylight}$. To this end, each such guess $\hat{\cal W}_{\ell, q}$ will be chosen as an integer multiple of $\frac{\eps^4 w_{\max}}{|\mathcal{Q}|}$, meaning that $\hat{\cal W}_{\ell, q} = \mu_{\ell, q} \cdot \frac{\eps^4 w_{\max}}{|\mathcal{Q}|}$ for some non-negative integer $\mu_{\ell,q}$.
Clearly, there exists a set of multiples $\{\mu_{\ell,q}\}_{\ell \in [L], q \in \mathcal{Q}_{\mylight}}$ such that $\mathcal{W}^{\uparrow}_{\ell, q} -\frac{\eps^4 w_{\max}}{|\mathcal{Q}|} \leq \mu_{\ell,q}\cdot\frac{\eps^4 w_{\max}}{|\mathcal{Q}|} \leq \mathcal{W}^{\uparrow}_{\ell, q}$ for every $\ell \in [L]$ and $q \in \mathcal{Q}_{\mylight}$. The important observation is that $\sum_{\ell \in [L]}\sum_{q \in \mathcal{Q}_{\mylight}}\mu_{\ell,q} \leq \frac{2|\mathcal{Q}|}{\eps^5}$, since 
\begin{eqnarray*}
    \sum_{\ell \in [L]}\sum_{q \in \mathcal{Q}_{\mylight}}\mu_{\ell,q} \cdot \frac{\eps^4 w_{\max}}{|\mathcal{Q}|} 
    &\leq & \sum_{\ell \in [L]}\sum_{q \in \mathcal{Q}_{\mylight}}\mathcal{W}^{\uparrow}_{\ell, q} \\
    & \leq & w(\mathcal{A}^{\uparrow}(\mathcal{B}^{\uparrow}_{[1, L]})) \\
    &< & \frac{2w_{\max}}{\eps} \ . 
\end{eqnarray*}
To better understand the last inequality, we first recall that each block $\mathcal{B}^{\uparrow}_{\ell}$ stretches up to and including position $p_\ell$, which is defined as the maximal position $p \in [n]$ for which $w(\mathcal{A}^{\uparrow}[1,p]) < (1+\eps)^{\ell} \cdot \eps^3 w_{\max}$. Specifically for block $\mathcal{B}^{\uparrow}_L$, since $L$ is defined in Section~\ref{subsec:block_bw} as the smallest index $\ell$ for which $(1+\eps)^{\ell}\cdot \eps^3 > \frac{1}{\eps}$, it follows that
\[w(\mathcal{A}^{\uparrow}(\mathcal{B}^{\uparrow}_{[1, L]})) ~~<~~ (1+\eps)^{L}\cdot \eps^3 w_{\max} ~~\leq~~  (1+ \eps)\cdot \frac{w_{\max}}{\eps} ~~\leq~~ \frac{2w_{\max}}{\eps} \ . \]

Consequently, the total number of joint guesses for $\{\mu_{\ell,q}\}_{\ell \in [L], q \in \mathcal{Q}_{\mylight}}$ is upper-bounded by the number of non-negative integer-valued solutions to $\sum_{\ell \in [L]}\sum_{q \in \mathcal{Q}_{\mylight}}\mu_{\ell,q} \leq \frac{2|\mathcal{Q}|}{\eps^5}$, which is precisely 
\begin{eqnarray}
    \binom{L \cdot |\mathcal{Q}_{\mylight}| + \frac{2|\mathcal{Q}|}{\eps^5}}{\frac{2|\mathcal{Q}|}{\eps^5}} & \leq & \binom{\frac{2|\mathcal{Q}|}{\eps^2}+\frac{2|\mathcal{Q}|}{\eps^5}}{\frac{2|\mathcal{Q}|}{\eps^5}} \label{w_eq_1} \\\nonumber\\
    & \leq &  2^{4|\mathcal{Q}|/{\eps^5}} \nonumber \\
    & = & O(n^{O(\frac{1}{\eps^6}\log \frac{1}{\eps})}) \ .  \label{w_eq_2}
\end{eqnarray}
Here, inequality~\eqref{w_eq_1} holds since $L \leq \frac{2}{\eps^2}$, as shown in our explanation for inequality~\eqref{tmid_3}. Equality~\eqref{w_eq_2} is obtained by plugging in $|\mathcal{Q}| = O(\frac{1}{\eps}\log(\frac{n}{\eps}))$. 

\subsection{Proof of Lemma~\ref{lem:clm_wbq_1}}\label{app_clm_wbq_1}

\paragraph{Bounded size.} We first show that $\mathcal{S} = (\mathcal{S}_1, \ldots, \mathcal{S}_L)$ satisfies property~\ref{good_assign_1}, stating that $|\mathcal{S}_{\ell}| \leq \beta^{\uparrow}_{\ell}$ for every $\ell \in [L]$. To this end, since $\beta^{\uparrow}_{\ell} = \sum_{q \in [Q]}\beta^{\uparrow}_{\ell,q}$, it suffices to show that $|\mathcal{S}_{\ell} \cap \mathcal{G}_q| \leq \beta^{\uparrow}_{\ell,q}$, for every $q \in [Q]$, and we proceed by considering two cases: 
\begin{itemize}
    \item {\em When $q \in \mathcal{Q}_{\myheavy}$}: In this case, it is worth pointing out once again that we focus on the scenario where  $\hat{\beta}_{\ell,q} = \beta^{\uparrow}_{\ell,q}$ for every $q \in \mathcal{Q}_{\myheavy}$. As explained in Section~\ref{sec_asgn_heavy}, exactly $\hat{\beta}_{\ell,q}$ products from $\mathcal{G}_q$ are assigned to $\mathcal{S}_{\ell}$, implying that $|\mathcal{S}_{\ell} \cap \mathcal{G}_q| = \hat{\beta}_{\ell,q} = \beta^{\uparrow}_{\ell,q}$.
    
    \item {\em When $q \in \mathcal{Q}_{\mylight}$}: Unlike heavy classes, it is quite possible that $|\mathcal{S}_{\ell} \cap \mathcal{G}_q|$ and $\beta^{\uparrow}_{\ell,q}$ may differ for light weight classes. However, as explained in Section~\ref{sec_asgn_light}, we assign exactly $\hat{n}_{\ell,q} \leq |\mathcal{A}^{\uparrow}(\mathcal{B}^{\uparrow}_{\ell}) \cap \mathcal{G}_q| = \beta^{\uparrow}_{\ell,q}$ products from $\mathcal{G}_q$ to $\mathcal{S}_{\ell}$, implying that $|\mathcal{S}_\ell\cap\mathcal{G}_q| \leq \beta^{\uparrow}_{\ell,q}$. 
\end{itemize}
Putting both bounds together, we conclude that
\[|\mathcal{S}_{\ell}|  ~~=~~  \sum_{q \in \mathcal{Q}_{\myheavy}}|\mathcal{S}_{\ell} \cap \mathcal{G}_q|  + \sum_{q \in \mathcal{Q}_{\mylight}}|\mathcal{S}_{\ell} \cap \mathcal{G}_q| 
     ~~\leq~~  \sum_{q \in [Q]} \beta^{\uparrow}_{\ell,q}
     ~~=~~  \beta^{\uparrow}_{\ell} \ . \]

\paragraph{Highest-index weight class.} We move on to proving that our partition meets property~\ref{good_assign_2}, which states that for every $\ell \in [L]$ with  $q^{\uparrow}_\ell \in {\cal Q}_{\myheavy}$, the set $\mathcal{S}_{\ell}$ contains at least one product from $\mathcal{G}_{q^{\uparrow}_\ell}$. Given that ${q}^{\uparrow}_{\ell} = \max\{q \in [Q]: \beta^{\uparrow}_{\ell, q} \geq 1\}$, our method of assigning $\beta^{\uparrow}_{\ell,q}$ products from each heavy weight class $\mathcal{G}_q$ to the set $\mathcal{S}_{\ell}$ indeed guarantees that $\mathcal{S}_{\ell} \cap \mathcal{G}_{q^{\uparrow}_{\ell}} \neq \emptyset$ when $q^{\uparrow}_{\ell} \in \mathcal{Q}_{\myheavy}$.

\paragraph{Prefix subsets.} Finally, we establish property~\ref{good_assign_4}, which states that the collection of products in $\mathcal{S}_{1}, \ldots, \mathcal{S}_L $ is a subset of those assigned by $\mathcal{A}^{\uparrow}$ to the blocks  $\mathcal{B}^{\uparrow}_0, \ldots, \mathcal{B}^{\uparrow}_L$. For this purpose, it suffices to show that $\mathcal{G}_q \cap (\bigcup_{\ell \in [L]} \mathcal{S}_{\ell}) $ is a subset of  $ \mathcal{G}_q \cap (\bigcup_{\ell \in [L]_0} \mathcal{A}^{\uparrow}(\mathcal{B}^{\uparrow}_{\ell}))$ for every $q \in [Q]$. 

We first recall that $|\mathcal{G}_q \cap \mathcal{S}_{\ell} | \leq \beta^{\uparrow}_{\ell,q}$ for every $\ell \in [L]$, as explained when proving property~\ref{good_assign_1} above. In addition, our assignment method ensures that ${\cal S}_1, \ldots, {\cal S}_L$ are collectively assigned the $\sum_{\ell \in [L]} |\mathcal{G}_q \cap \mathcal{S}_{\ell}|$-lowest ranked products in the sequence $\langle 1, q \rangle, \ldots, \langle |\mathcal{G}_q|, q \rangle$, which is an ordering of $\mathcal{G}_q$. These elements form a prefix of the $\sum_{\ell \in [L]_0} \beta^{\uparrow}_{\ell,q}$-lowest ranked products, which are precisely those assigned by $\mathcal{A}^{\uparrow}$ to the collection of blocks $\mathcal{B}^{\uparrow}_0, \ldots, \mathcal{B}^{\uparrow}_L$. Therefore, $\mathcal{G}_q \cap (\bigcup_{\ell \in [L]} \mathcal{S}_{\ell}) $ is indeed a subset of  $ \mathcal{G}_q \cap (\bigcup_{\ell \in [L]_0} \mathcal{A}^{\uparrow}(\mathcal{B}^{\uparrow}_{\ell}))$.